\documentclass[11pt,a4paper]{article}

\usepackage{jheppub}
\usepackage{caption, subcaption}  

\def\beq {\begin{equation}}
\def\eeq {\end{equation}}
\def\bi {\begin{itemize}}
\def\ei {\end{itemize}}
\def\bea {\begin{eqnarray}}
\def\eea {\end{eqnarray}}

\numberwithin{equation}{section} 

\title{Study of energy deposition patterns in hadron calorimeter for prompt and displaced jets using convolutional neural network}

\author[a]{Biplob Bhattacherjee}
\author[b]{Swagata Mukherjee}
\author[a]{Rhitaja Sengupta}

\affiliation[a]{Centre for High Energy Physics, Indian Institute of Science,\\
Bangalore 560012, India}

\affiliation[b]{III. Physikalisches Institut A, RWTH Aachen University,\\
Otto-Blumenthal-Str. 16, 52074 Aachen, Germany}

\emailAdd{biplob@iisc.ac.in}
\emailAdd{mukherjee@physik.rwth-aachen.de}
\emailAdd{rhitaja@iisc.ac.in}


\abstract{
Sophisticated machine learning techniques
have promising potential in search for physics beyond Standard Model in 
Large Hadron Collider (LHC). 
Convolutional neural networks (CNN)
can provide powerful tools for differentiating between patterns of calorimeter energy deposits by prompt particles of Standard Model and long-lived particles predicted in various models beyond  
the Standard Model. We demonstrate the usefulness of CNN by using a couple of physics examples from well motivated BSM scenarios predicting long-lived particles giving rise to displaced jets.
Our work suggests that modern machine-learning techniques 
have potential to discriminate between energy deposition patterns of prompt and long-lived particles, and thus, they can be useful tools in such searches.}

\begin{document}
\maketitle
\flushbottom
	

    



\section{Introduction}
\label{sec:intro}

The Standard model (SM) of particle physics is an extremely successful model in describing the interactions between known elementary particles. 
However, it cannot be a complete theory, because it does not address 
several open questions as well as observations in particle physics, such as neutrino oscillations, matter-antimatter asymmetry, dark matter, etc. 
This implies that new particles beyond the Standard Model (BSM) should exist.
Current BSM search programs in LHC experiments mostly focus on energetic final states produced
at the interaction point (IP). 
However, the existing searches are largely insensitive to Long-Lived Particles (LLPs), which decay into SM 
particles some macroscopic distance away from the IP.
Particles with macroscopic lifetimes
are ubiquitous in the SM, and the same mechanisms could be present in any BSM theory.
Theories with LLPs, such as many versions of hidden sectors or hidden valleys~\cite{Strassler:2006im, Curtin:2014cca, Liu:2018wte, Barman:2018pez, Alipour-Fard:2018lsf}, 
SUSY~\cite{Giudice:1998bp,Farrar:1996rg,Baer:1998pg,Mafi:1999dg,Kraan:2004tz,ArkaniHamed:2004fb,Giudice:2004tc,Hewett:2004nw,Arvanitaki:2005nq,Meade:2010ji, Fan:2012jf, Bhattacherjee:2012ed, Arvanitaki:2012ps, Banerjee:2016uyt, Nagata:2017gci, Banerjee:2018uut, Ito:2018asa}, WIMP Baryogenesis~\cite{Cui:2014twa, Choi:2018kto}, neutrino extensions~\cite{Mohapatra:1974gc, Dev:2017dui, Cottin:2018kmq, Dev:2018kpa} etc., can address some or all of the fundamental shortcomings of the SM. Despite the strong motivations, LLPs mostly remain unexplored by standard LHC searches, although in recent years these particles have attracted growing
interest, and some dedicated searches have been performed in LHC experiments~\cite{Sirunyan:2018vlw, Sirunyan:2018njd, Sirunyan:2018pwn, Sirunyan:2017sbs, Sirunyan:2018ldc, Aaboud:2019opc, Aaboud:2019trc, Aaboud:2018kbe, Aaboud:2018aqj, Aaboud:2018arf, Aaij:2017mic, Aaij:2016xmb, Alimena:2019zri}.\footnote{A comprehensive overview of LLP signatures at LHC and recent searches can be found in \cite{Alimena:2019zri} and the references therein.}

LHC searches depend on reconstruction of objects, like photons, electrons, muons and jets, in an event. These reconstructions are based on standard algorithms taking information from different parts of the detectors. In collider experiments, the calorimeters (both electromagnetic, ECAL and hadronic, HCAL) are segmented in $\eta-\phi$ towers in such a way that a single particle coming from the primary vertex usually falls within a single calorimeter tower corresponding to the particle's actual $\eta$ and $\phi$.
For LLPs, where the particles have long lifetimes and therefore decay after travelling sufficiently longer distances, the situation may differ. If the decay products are displaced enough, then their $\eta$ and $\phi$ directions (which start from the secondary vertex) may not match with the standard $\eta-\phi$ segmentation of the detectors. 
Therefore 
a particle is no longer guaranteed to be concentrated to a single $\eta-\phi$ tower. Rather, it may be spread over several $\eta-\phi$ towers (as shown in Fig.~\ref{fig:segment}).~
This feature was studied for displaced photons which look quite different in the ECAL than the prompt ones \cite{Chatrchyan:2012jwg}.
One might then expect that this effect 
will be more prominent for displaced jets which contain many particles; and the energy deposition in the HCAL should be different from standard patterns observed for prompt jets. 
However, it may be the case that many particles actually smear out this effect, or, that we don't see such effects due to the coarser resolution of the HCAL, compared to ECAL.
Therefore we attempt to study how displaced jets look in the HCAL and  understand their different features.


Long lived particles which are color-neutral will decay to two or more quarks or gluons, and therefore, we will 
have more than one displaced jet in the final state. If the LLP has color charge, it will most probably hadronise before decaying (when the QCD phase-transition scale is greater than the decay width of the LLP, i.e., $\Lambda_{QCD}>\Gamma$). These hadrons are again color neutral objects which later decay to give multiple jets. Therefore, we mostly land up with 
at least two displaced jets in the final state.
Standard displaced jets analysis of ATLAS and CMS might lose its sensitivity as the distance between the IP and the secondary vertex increases, a probable reason being that the displaced jets will have different energy deposition patterns in the HCAL compared to the standard 
pattern we usually see for prompt jets. This makes the reconstruction of these displaced jets 
challenging as the decay length of the LLP increases. Even with a few tweaks in the reconstruction algorithm, this task is quite arduous. It may be the case that in present experiments, we mostly lose such LLP events with standard 
reconstruction techniques, and thus end up with low signal efficiency. Typical signal efficiency in analyses involving displaced jets signature in LHC experiments is often in the ballpark of 
a few percent~\cite{Sirunyan:2018vlw}. 
Another experimental challenge that leads to reduced signal efficiency is triggering. ATLAS and CMS collaborations have come up with several novel ideas to 
overcome challenges with triggering LLP signatures. One such idea is to use ``CalRatio'' trigger~\cite{Aaboud:2019opc, Aad:2013txa}, which is effective to identify jets that result from LLPs 
decaying near the outer radius of the ECAL or within
the HCAL, i.e., they will not have enough ECAL deposit as prompt jets would have. This triggering process however would not work well if the decay of LLP takes place near the outer tracker, which is the case mostly discussed in this paper.

Another novel idea, by the ATLAS collaboration, is to trigger on trackless jet signature, which provides a powerful handle for identifying displaced jets~\cite{Aad:2013txa}. However, it was pointed out in the paper by ATLAS collaboration that the trackless jet trigger is less than $5\%$ efficient for decays occurring in the region of the
barrel between the middle of the inner tracking system and the ECAL. The trigger efficiency increases with the decay length and is maximum where the decay occurs closer to the end of the HCAL. This calls for the need to look for new ideas and techniques which can serve as additional search strategies for LLPs. Going beyond standard cut-based approach seems necessary, as standard cuts neglect possible correlations between variables
which can be used to better identify signals. 
In this work we focus only on HCAL energy deposits to distinguish between prompt jet and displaced jet. In real data analysis in experiments, it is possible to use our proposed method together with information from other sub-detectors. For example, where LLP decays inside tracker, it is possible to make use of displaced track information. But one needs to keep in mind that reconstruction of displaced track comes with a non-negligible amount of inefficiency, as shown in ~\cite{Chatrchyan:2014fea}~(figure 12 in particular), and high misidentification rate.
Since reconstruction efficiency of displaced objects tend to be smaller than their prompt counterpart, in this work, we take a step back from reconstructing jets to just studying their energy deposition patterns in the HCAL. For that we have simulated a toy calorimeter closely resembling the barrel HCAL of the ATLAS detector.
We consider two 
scenarios. First, displaced SM $Z$ boson coming from the decay of an LLP and further decay of $Z$ to give displaced jets and second, displaced jets directly coming from the decay of an LLP. These two scenarios have some kinematic differences $-$ the former being a two-body decay of the LLP and subsequent decay of an on-shell SM $Z$ boson to jets and the latter being a three-body decay of the LLP. We also make both the scenarios boosted enough to bring all the LLP decay products closer such that they can be viewed together 
in the calorimeter. We consider distributions of energy fraction deposited in $\eta-\phi$ regions of various sizes and find these distributions to have some differences for different displacements of the LLP. This motivates us to look for modern ways of classifying LLP and non-LLP scenarios using the difference in energy deposition pattern.

Machine learning 
has recently shown great results in many disciplines and has a variety of applications. It offers the potential to automate challenging data-processing tasks, and classification in high-dimensional variable spaces in collider physics~\cite{Albertsson:2018maf}. Neutrino experiments like NOvA~\cite{Adamson:2017zcg} and MicroBooNE~\cite{Acciarri:2016ryt} used computer-vision techniques to reconstruct and classify various types of neutrino events. Our approach builds upon the paradigm that a substantial energy deposit in HCAL, coming from a SM particle or from a BSM particle, can be treated as an image, with intensity given by the amount of energy deposit in HCAL. One straight-forward approach to process the LHC data is the jet images approach discussed  in \cite{Cogan:2014oua, Almeida:2015jua, deOliveira:2015xxd}. The idea behind jet images is to treat the energy deposits in a calorimeter as intensities in a 2D image. Then one can apply sophisticated algorithms developed for image recognition. This and related neural network approaches were used for boosted $W$ boson tagging~\cite{Cogan:2014oua,deOliveira:2015xxd,Baldi:2016fql,Larkoski:2017jix}, quark-gluon discrimination \cite{Komiske:2016rsd,ATL-PHYS-PUB-2017-017,Chien:2018dfn}, top tagging~\cite{Pearkes:2017hku, Almeida:2015jua, Kasieczka:2017nvn, Kasieczka:2019dbj}, generic anti-QCD tagging~\cite{Aguilar-Saavedra:2017rzt, Roy:2019jae}, photon identification \cite{Ghosh:2018gyw}, heavy-flavor tagging~\cite{Guest:2016iqz} and search for BSM particles~\cite{Farina:2018fyg, Cerri:2018anq, Collins:2019jip, Hajer:2018kqm, Heimel:2018mkt, Chakraborty:2019imr}. 
Modern deep learning algorithms trained on HCAL energy deposition images can be used for LLP studies. We use convolutional neural network (CNN), a supervised learning algorithm which is specifically designed for working with images. We train a CNN to identify the features associated with displaced jets  
so that it can distinguish the LLP events from non-LLP 
ones based on the images of energy deposition patterns in the HCAL.

To the best of our knowledge, this work is the first attempt in studying LLPs using image recognition algorithms like convolutional neural networks. We have used minimal preprocessing to the images in order to be as model-independent as possible. We have not done advanced optimisations in this work since our goal is just to understand the differences in energy deposition due to displacement and further optimisations might achieve better discrimination. 
This paper is outlined as follows: in section \ref{ssec:segment_calo}, we present the segmentation of the HCAL used by us; in section \ref{ssec:disp_energy}, we discuss the two displaced scenarios: displaced $Z$ boson and displaced jets coming directly from the decay of an LLP. We also discuss the change in pattern of 
energy 
deposition with displacement using distributions of energy fractions for the two scenarios and identify the key features associated with displaced multijet systems; in section \ref{ssec:cnn}, we describe the CNN architecture used by us and other details related to the network training, validation and testing; the analysis and results are presented in section \ref{ssec:analysis} where we check the effect of mass of the LLP in the displaced $Z$ scenario 
 and also present a brief discussion on stopped particle scenario as a special case of the second displaced scenario described above, in which the LLP stops in the HCAL and then decays to give jets; finally in section \ref{sec:concl}, we conclude.  

\section{Understanding the displaced features of LLPs using images}
\label{sec:full}

The main motive of this work is to study whether displaced jets coming from color neutral LLPs have any difference in energy deposition patterns in the HCAL from standard prompt jets. Furthermore,  if there is any difference, we need to know whether this difference is significant enough so that ordinary analyses with cuts can distinguish between them or does it call for the need to employ image recognition techniques for such discrimination. These differences may also depend on various LLP models and mass of the LLP. In this work we aim to address such questions. We will identify the features associated with only the displacement of jets and therefore, compare the non-displaced and displaced cases of the same model. We will do the same for both the scenarios which are described in section \ref{sec:intro} to find out how the difference in these two scenarios highlight slightly different features of displaced multijet systems.

For observing features associated with displacement of a particle, we have to consider the actual segmentation of the calorimeter. The usual analysis with fast detector simulation (for example, \texttt{Delphes}) which only has $\eta-\phi$ segmentation of the calorimeters won't work here because it has no layered calorimeter structure and no segmentation in the physical $z$ direction. 
Such detector simulations give the actual $\eta-\phi$ of the LLP decay products. As a result, we don't get the effects of displaced particles as discussed in section \ref{sec:intro} which arises due to projection of radial layers taken along constant $\eta-\phi$ directions and also due to segmentation in $z$ direction.
Therefore, for the illustration of our idea, we simulate a simplified calorimeter with segmentation close to that of the ATLAS HCAL. We start by discussing the segmentation of our toy calorimeter in the next section.

\subsection{Segmentation of the calorimeter}
\label{ssec:segment_calo}

In this section, we explain the segmentation of the HCAL that we have considered here and also elucidate how we propagate the particles and parametrise their energy deposition in each tower of the calorimeter.

\begin{figure}[hbt!]
\centering
\includegraphics[width=15cm]{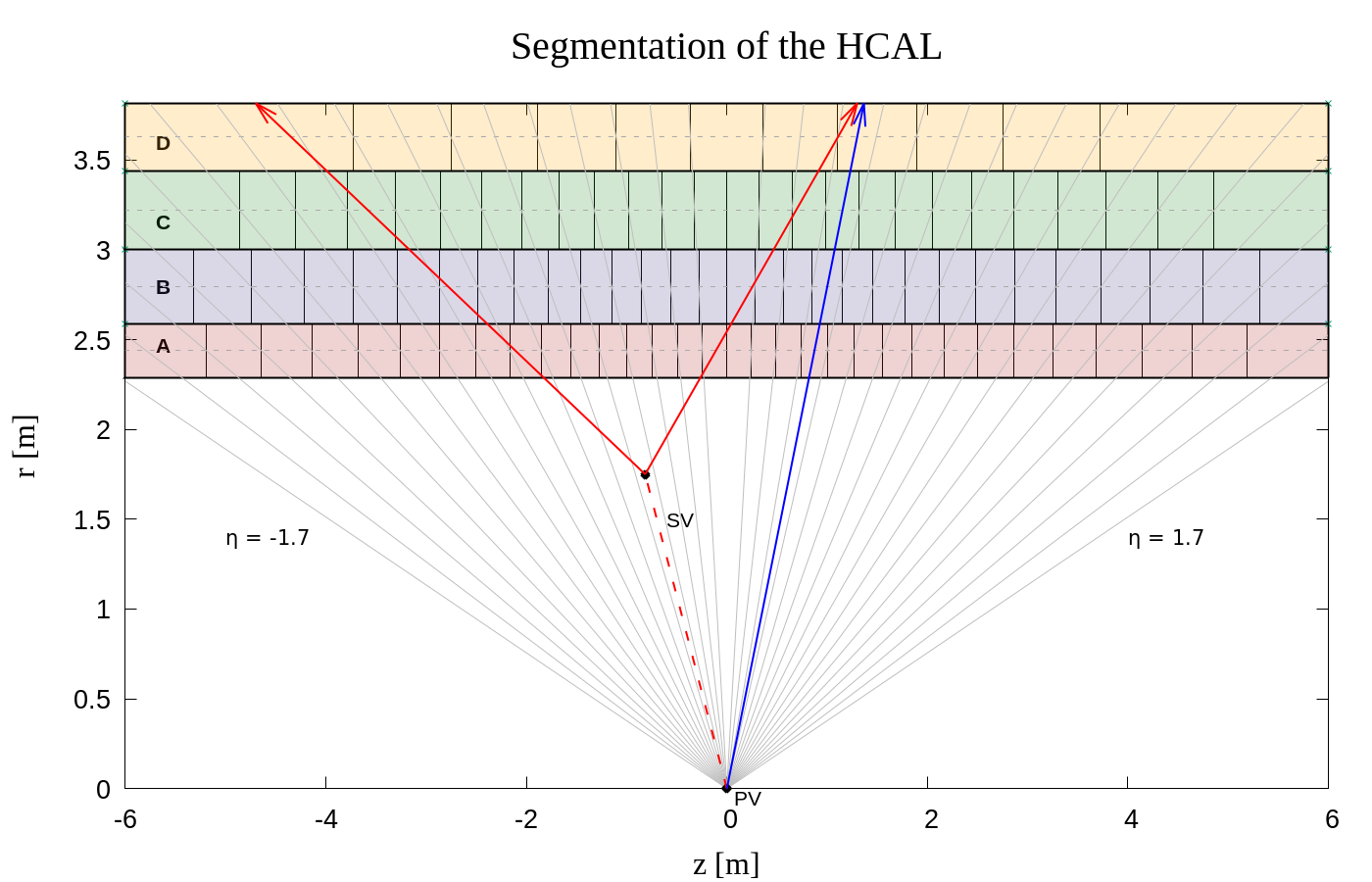}
\caption{Simplified segmentation of the HCAL used in this work $-$ A,B,C and D are the four radial layers extending from 2.2895 m to 3.8158 m and have a pseudorapidity coverage of $|\eta| < 1.7$. Constant $\eta$ lines are shown in gray for reference. A standard particle coming from the Primary Vertex (PV) is shown in blue. A long-lived particle coming from PV and its decay products starting from the secondary vertex (SV) are shown in red. Note that the particle in blue is contained within a single $\eta$ tower (between two grey lines) whereas the particles in red are spread over several $\eta$ towers.}
\label{fig:segment}
\end{figure}

For the segmentation part, we closely follow the Tile Calorimeter~\cite{Aad:2010af} of the ATLAS detector which has four layers, viz., A, B, C and D, arranged along the radial direction, one after the other, with segmentations in the $\eta-\phi$ directions. The $\phi$ direction is segmented into $64$ parts $-$ each segment being close to $0.1$ radian. For segmentation in the $\eta$ direction, we take the segment size to be $0.1$ radian for layers A, B and C, and $0.2$ radian for layer D, the same as ATLAS. Now 
to translate this $\eta$ segmentation into $z$ segmentation,  
we 
segment along the $z$ points 
where the constant $\eta$ lines (0.1 radian apart) cut the radial central line of each layer $-$ same as done in the Tile Calorimeter. We have shown 
in Fig.~\ref{fig:segment}~the segmentation of the calorimeter along the $z$ direction with constant $\eta$ lines for reference. 
In the $z$-direction, we extend upto $|z|=6$ m which corresponds to the barrel region of the ATLAS HCAL detector. This gives a pseudorapidity coverage of about $|\eta| < 1.7$.\footnote{The pseudorapidity coverage of the calorimeter can be extended by following the segmentation of the end-cap parts of the Tile Calorimeter. In this work, we restrict to the barrel region of the calorimeter.}

We now generate events of physical processes and use the \texttt{ParticlePropagator} code in \texttt{Delphes-3.4.1} \cite{deFavereau:2013fsa} to propagate the particles to the starting of layer A of the HCAL.  
If the particle has $|z| > 6 {\rm~m}$ already by then, we ignore it. We also ignore electrons, muons and neutrinos. We then propagate each particle further to find out the $\eta-\phi$ bins of the HCAL traversed by the particle 
 and store the energy deposited by the particle in each of these bins. The energy deposited in each bin is 
 directly proportional to the distance travelled by the particle in that bin. This is done for all the particles in an event and then projection is taken along constant $\eta$ and $\phi$ directions of all the four layers. This gives us energy deposition in the $\eta-\phi$ plane which is divided into $32\times64$ towers. A validation of our HCAL segmentation with \texttt{Delphes} fast simulation for prompt jets is presented in appendix \ref{app:val}.

Any tower of the HCAL having energy deposit less than $1 {\rm~GeV}$ is ignored. We normalise the energy in each tower of an event using the maximum \footnote{We have also normalised the images with the energy sum of each event and repeated the analysis. This has negligible effect on our results.}~energy deposited in the HCAL for that event.
We store the energy deposition of an event as a $28$ pixel$\times28$ pixel image with the energy depositions in each tower as intensity values of each pixel of the image with the 
highest intensity (energy) pixel at the centre of the $i\eta-i\phi$\footnote{$i\eta$ and $i\phi$ implies the bin number in the $\eta-\phi$ directions respectively \cite{Chadeeva_2018}. We use this notation to denote pixels of the images.}~plane, i.e., the (14,14) pixel. The reason why we have considered images of size $28$ pixel$\times28$ pixel is that the jets are boosted enough (in the energy range that we are considering in this work) such that $\Delta R$ between any two jets is always less than $1.4$ and hence, the energy deposition of the multijet system in each case is contained within $28\times28$ region of $\eta$-$\phi$ plane. We use these images as our input to the neural network.

\subsection{Two displaced scenarios and their energy deposition patterns}
\label{ssec:disp_energy}

In this section, we describe the two displaced scenarios considered by us for this work and their simulation details.
We also discuss the difference in their energy deposition patterns in the HCAL and how and why this pattern changes with displacement.

Using the segmentation process described in section \ref{ssec:segment_calo}, we now have $28\times28$ images of energy deposition in the calorimeter for the following four cases:
\begin{list}{$\bullet$}{}
\item particles from the PV, hence, not displaced;
\item particles from a transverse distance between $30 {\rm~cm}$ and $50 {\rm~cm}$;
\item particles from a transverse distance between $50 {\rm~cm}$ and $70 {\rm~cm}$;
\item particles from a transverse distance between $70 {\rm~cm}$ and $90 {\rm~cm}$;
\item particles from a transverse distance between $200 {\rm~cm}$ and $220 {\rm~cm}$,
\label{list:disp_cases}
\end{list}

where the first four are cases of LLP decay within the Tracker and the last one corresponds to decay of the LLP just before entering the HCAL. Hereafter, whenever we mention displacement of particles, we mean displacement in the transverse direction, unless otherwise stated.

To study the difference in energy deposition pattern with displacement, we consider the following distributions:
$$f_i,~~~i = 3, 5, 9, 11$$
where $f_i$ is the fraction of energy deposited in $i\times i$ block of the full $28\times28$ image. This $i\times i$ block's centre is anchored to the centre of the $28\times28$ image which is the highest energy pixel.

\begin{figure}[hbt!]
\centering
\includegraphics[scale=0.245]{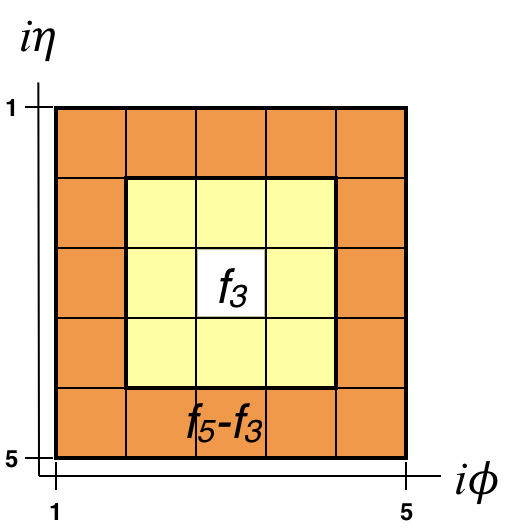}
\caption{Central $5\times5$ part of the $28\times28$ $i\eta-i\phi$ image. The fraction of total energy (in $28\times28$ $i\eta-i\phi$ region) deposited in this part is $f_5$. The central $3\times3$ part has energy fraction $f_3$ (yellow and white) and the rest has energy fraction $f_5-f_3$ (orange). $f_3$ contains the central highest energy tower (shown in white).}
\label{fig:energy_fraction}
\end{figure}

Fig. \ref{fig:energy_fraction} illustrates the meaning of $f_3$ and $f_5-f_3$. We will consider distributions of the energy fraction $f_3$ which includes the central highest energy tower; and exclusive energy fractions $-$ energy deposition fraction in $i\times i$ block excluding the previous block's energy fraction (like $f_5-f_3$, $f_9-f_5$, $f_{11}-f_9$ and so on \footnote{We have also studied distributions of exclusive energy fractions involving $f_{15}$, $f_{21}$ and $f_{28}$. These exclusive regions have very little energy deposits for most events and follow the same trend as $f_{11}-f_9$.}) to study differences in energy deposition patterns with displacement.

\subsubsection{Displaced jets from displaced SM $Z$ boson}  
\label{sssec:disp_Z}

The first 
scenario considered in this work is displaced jets 
from 
the decay of a boosted $Z$ boson, which again comes from the decay of a long-lived particle $X$. Hence, the $Z$ boson is displaced. From here on, we will refer to the LLP as $X$ throughout the subsequent sections. We consider that the $X$ decays to SM $Z$ boson and any invisible particle $Y$.
$$X{\rm~(LLP)} \rightarrow Z{\rm~(SM)}+Y{\rm~(Invisible)}$$

To simulate such a process, we use the gauge mediated SUSY breaking model~\cite{Kolda:1997wt,Giudice:1998bp}, where the neutralino is long-lived and decays to SM $Z$ boson and light gravitino which is invisible to the detector.
We consider the case where the $Z$ boson decays to quarks, $Z \rightarrow q \bar{q}$ and 
 are interested to check the pictorial view of energy deposition of this $Z \rightarrow q \bar{q}$, where the $Z$ is displaced since it's coming from the LLP decay, in HCAL in $\eta-\phi$ plane.

We use \texttt{PYTHIA6}~\cite{Sjostrand:2000wi} for generation of events. We consider a non-displaced $Z$ scenario in which the particle $X$ is not long-lived and therefore decays at the PV and displaced $Z$ scenario (coming from long-lived $X$) which again we divide in four categories as described in the starting of section \ref{list:disp_cases}. Also, to ensure that the decay products fall mostly within the calorimeter segment, we put a condition that the longitudinal displacement is less than $200 {\rm~cm}$. For comparing the different displaced and non-displaced cases, we apply an energy window cut and restrict to events where the total HCAL deposit inside the $28\times28$ part of the $i\eta-i\phi$ plane is between $(400,500) {\rm~GeV}$ for the non-displaced as well as all the displaced cases. The choice of this particular energy range is made to have boosted $Z$ bosons. Boost is required to bring the displaced jets closer in the $\eta-\phi$ plane.

\begin{figure}[hbt!]
\centering
\begin{subfigure}{0.5\textwidth}
\centering
\includegraphics[width=7.5cm]{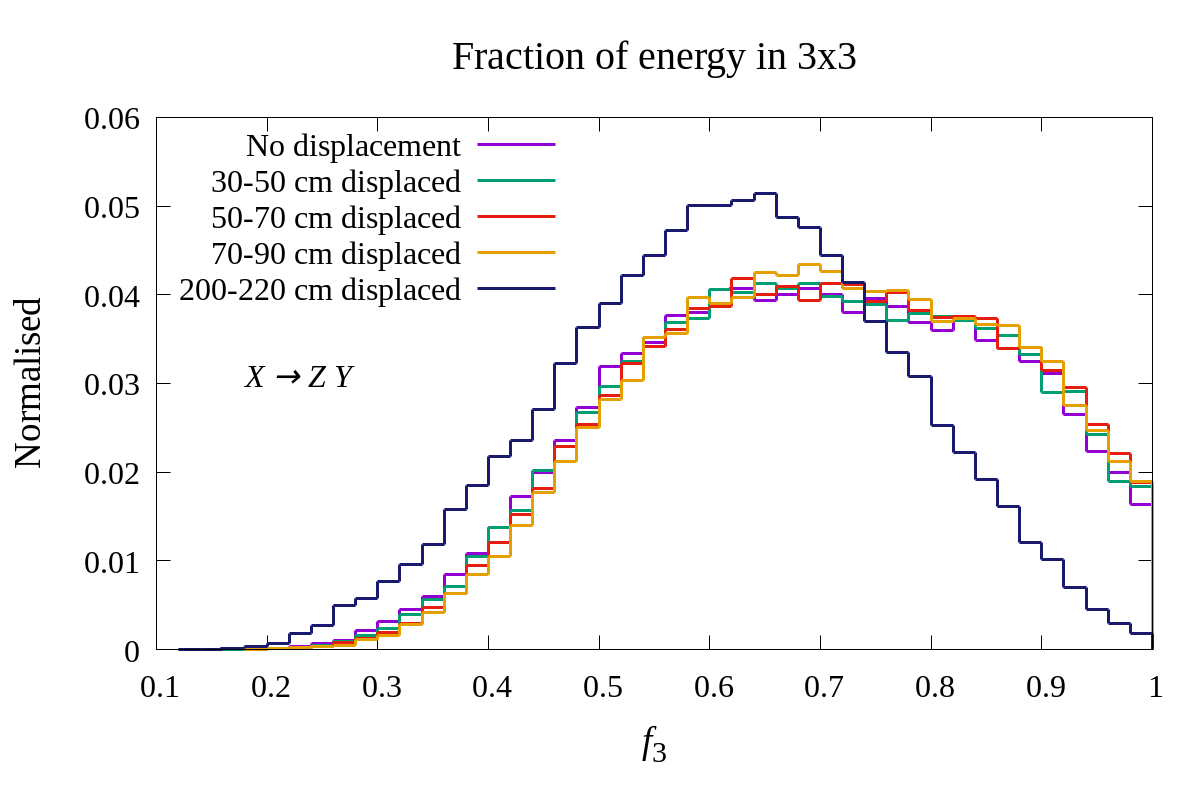}
\caption{}
\label{fig:3x3_gmsb}
\end{subfigure}%
\begin{subfigure}{0.5\textwidth}
\centering
\includegraphics[width=7.5cm]{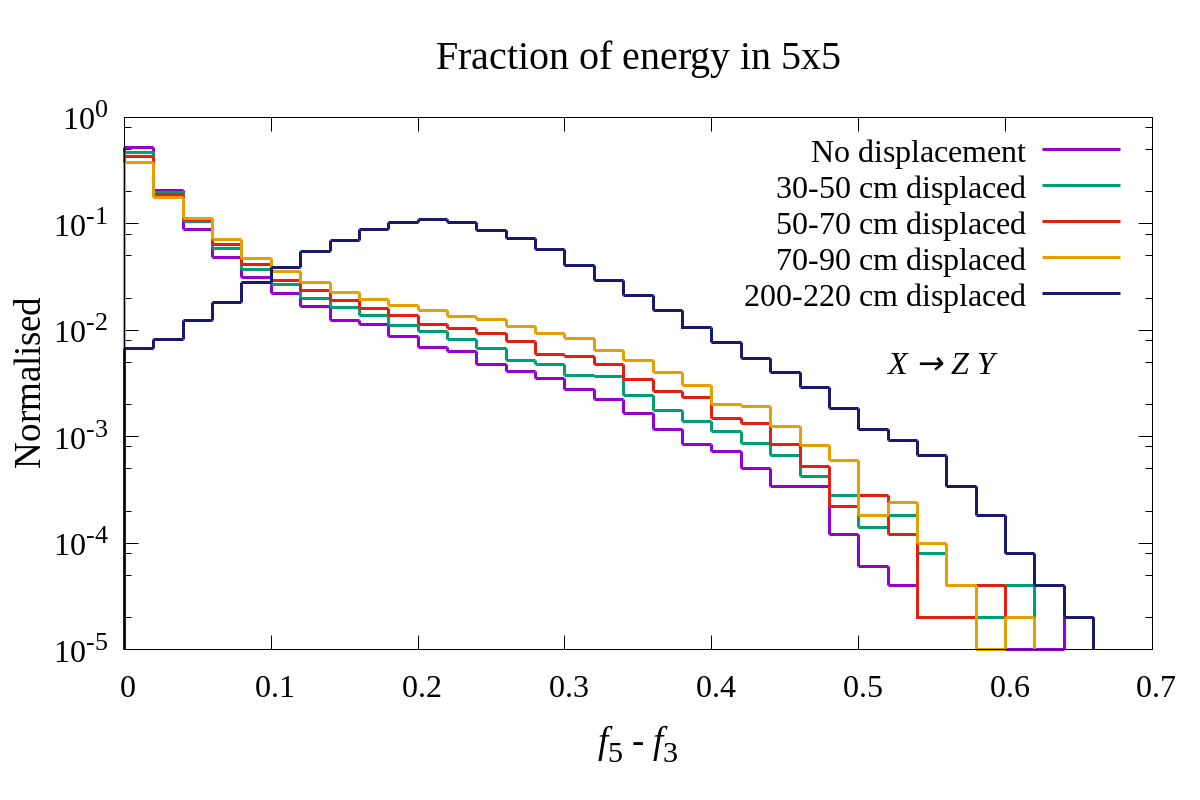}
\caption{}
\label{fig:5x5_gmsb}
\end{subfigure}\\
\begin{subfigure}{0.5\textwidth}
\centering
\includegraphics[width=7.5cm]{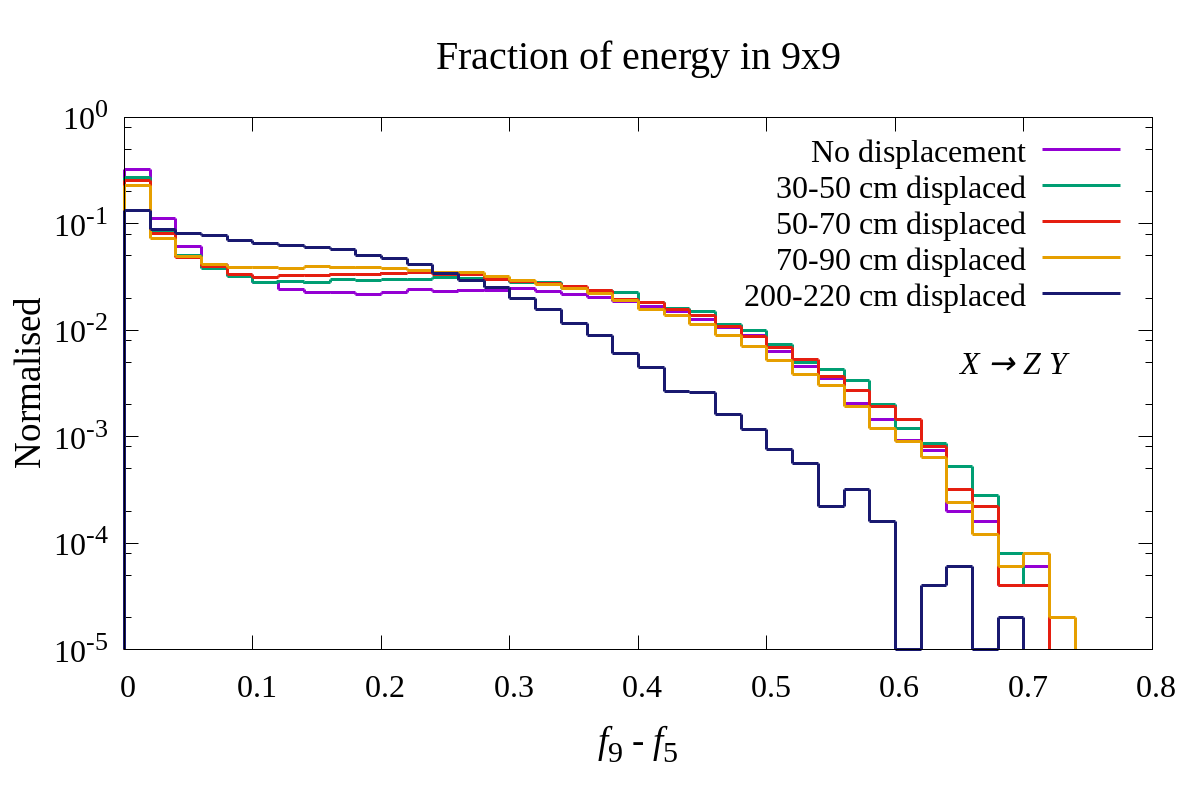}
\caption{}
\label{fig:9x9_gmsb}
\end{subfigure}%
\begin{subfigure}{0.5\textwidth}
\centering
\includegraphics[width=7.5cm]{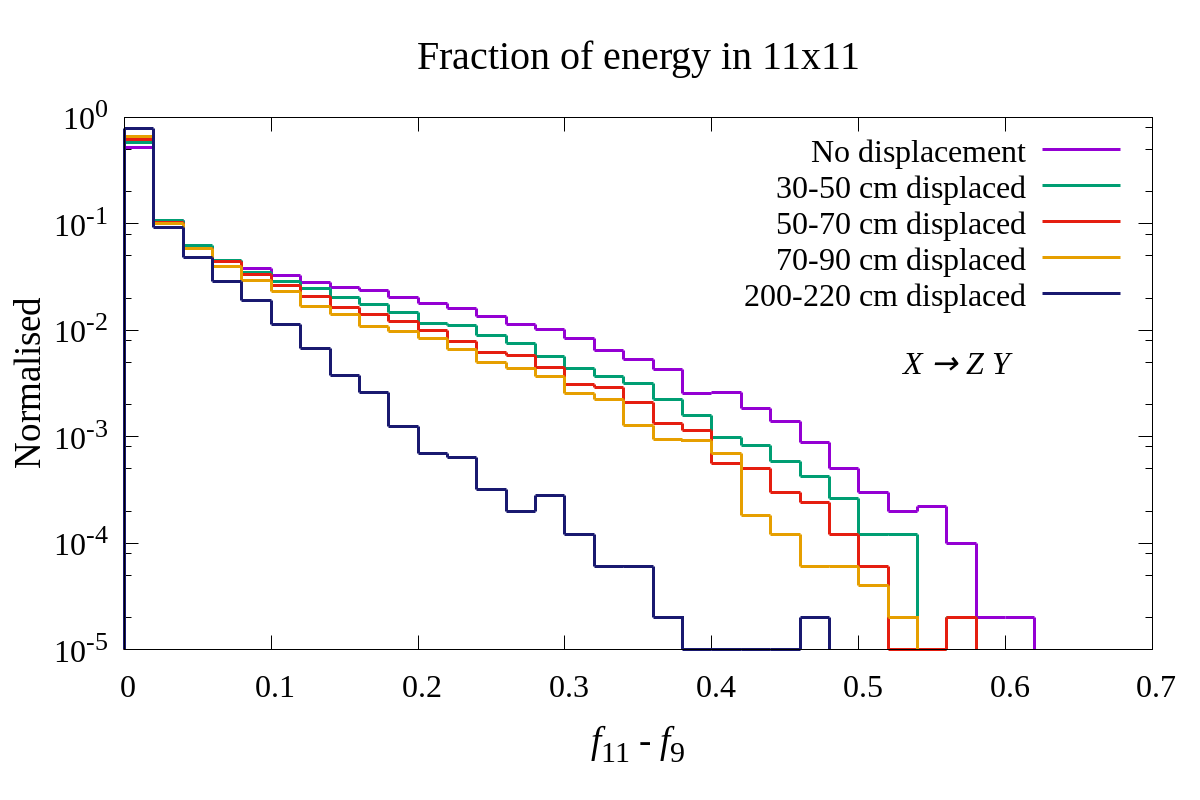}
\caption{}
\label{fig:11x11_gmsb}
\end{subfigure}%
\caption{Normalised distributions of energy deposition fraction ($f_i$) with varying sizes of blocks ($i\times i$). (a) Fraction of energy deposited in $3\times3$; (b),(c),(d) Fraction of energy deposited in $5\times5$, $9\times9$ and  $11\times11$ excluding the previous $3\times3$, $5\times5$ and $9\times9$ blocks' energy deposition fraction respectively.}
\label{fig:energy_gmsb}
\end{figure}

Fig.~\ref{fig:energy_gmsb} shows the normalised distributions of energy fractions for the non-displaced and the different displaced cases discussed above. Fig.~\ref{fig:3x3_gmsb} shows the energy fraction deposited in $3\times3$ block of the full image, $f_3$. To understand the feature shown by these distributions, let's first observe the non-displaced case and the first three displaced cases where $X$ decays within the tracker. There are only small differences in the distributions of $f_3$ with displacement for these cases. As we move to the energy deposited after this $3\times3$ block but within a $5\times5$ block, $f_5-f_3$, we note in Fig.~\ref{fig:5x5_gmsb} that the non-displaced $Z$ has relatively smaller energy deposition fraction in this region than the displaced $Z$ cases. In the next $9\times9$ block (exclusive of the $5\times5$), $f_9-f_5$, we see from Fig.~\ref{fig:9x9_gmsb} that the non-displaced $Z$ distribution has caught up with the displaced cases and again the difference is small. But for the next $11\times11$ block, $f_{11}-f_9$, we find that the non-displaced $Z$ has higher energy fractions than the displaced cases (Fig.~\ref{fig:11x11_gmsb}).

The change in these distributions corresponds to the fact that there are two distinct objects present in the decay of the $Z$ boson and there is a finite gap between these two jets in the $\eta-\phi$ plane. The distribution of $f_5-f_3$ is falling faster than the $f_9-f_5$ distribution and the latter has a small 
hump around $30-40\%$ energy fraction, implying that the second decay product mostly lies within the region between $5\times5$ and $9\times9$ in the $i\eta-i\phi$ plane. For the displaced $Z$ boson these objects are spread over the $\eta-\phi$ plane and there is a chance of overlap. Greater the displacement, greater is the chance of such an overlap and therefore the gap is not prominent for displaced cases. This explains the fact that the $f_5-f_3$ distribution falls off faster for the prompt $Z$ than the $70-90 {\rm~cm}$ displaced $Z$. Also, the $f_{11}-f_9$ distribution falls much faster for the displaced cases, implying that their total energy deposit is mostly contained within smaller $i\eta-i\phi$ region. The energy deposit of the second object shifts to more smaller $i\eta-i\phi$ blocks with increasing displacement. Looking at the distributions for the $200-220 {\rm~cm}$ displaced $Z$, we find that 
the containment of the total energy deposit within smaller $i\eta-i\phi$ regions is more prominent here and the second object mostly lies between the $3\times3$ and $5\times5$ blocks as is evident from the peak in $f_5-f_3$ distribution.

\begin{figure}[hbt!]
\centering
\begin{subfigure}{0.5\textwidth}
\centering
\includegraphics[width=7.5cm]{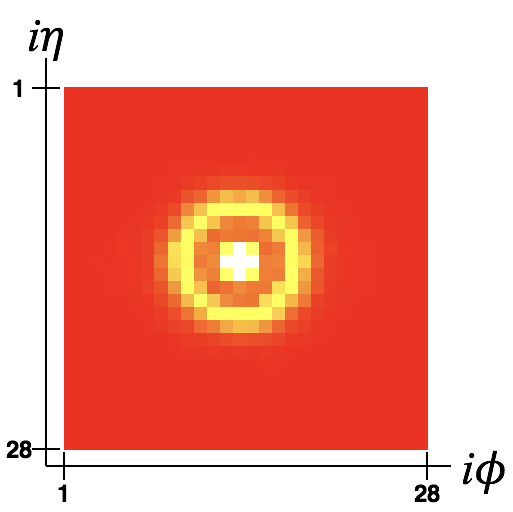}
\caption{}
\label{fig:0disp_gmsb}
\end{subfigure}%
\begin{subfigure}{0.5\textwidth}
\centering
\includegraphics[width=7.5cm]{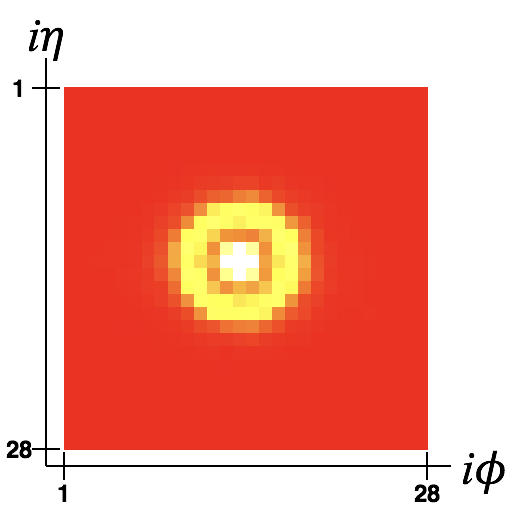}
\caption{}
\label{fig:30-50_gmsb}
\end{subfigure}\\
\begin{subfigure}{0.5\textwidth}
\centering
\includegraphics[width=7.5cm]{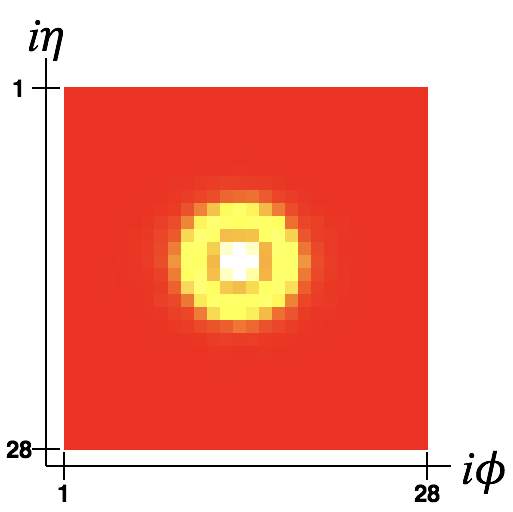}
\caption{}
\label{fig:50-70_gmsb}
\end{subfigure}%
\begin{subfigure}{0.5\textwidth}
\centering
\includegraphics[width=7.5cm]{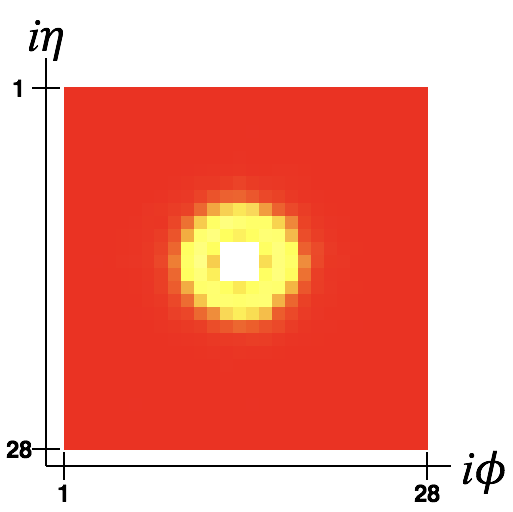}
\caption{}
\label{fig:70-90_gmsb}
\end{subfigure}%
\caption{Average of 50,000 images for (a) $Z$ boson with no displacement; (b), (c), (d) $Z$ boson displaced within transverse distance of $30-50 {\rm~cm}$, $50-70 {\rm~cm}$ and $70-90 {\rm~cm}$ from PV respectively. These are $28\times28$ images in the $i\eta-i\phi$ plane with the highest energy bin in the center (14,14). With increasing displacement, the distance between two jets decreases, as seen from the figures.}
\label{fig:avg_gmsb}
\end{figure}

Fig.~\ref{fig:avg_gmsb} shows the average of 50,000 $Z$ boson images for each of the cases where the decay is within the Tracker. For the non-displaced $Z$ (Fig.~\ref{fig:0disp_gmsb}), there is a clear annulus of the $Z$ boson coming from the second decay product. We observe that statistically these energy deposits look different. The distance between the two energy deposits corresponding to the two jets keeps decreasing with displacement and the overall size of the total energy deposit also becomes slightly smaller with increasing displacement. We have explained these displacement features at the end of section \ref{sssec:disp_chi} after discussing the energy deposition pattern of the second scenario.

\begin{figure}[hbt!]
\centering
\begin{subfigure}{0.49\textwidth}
\centering
\includegraphics[width=0.95\textwidth]{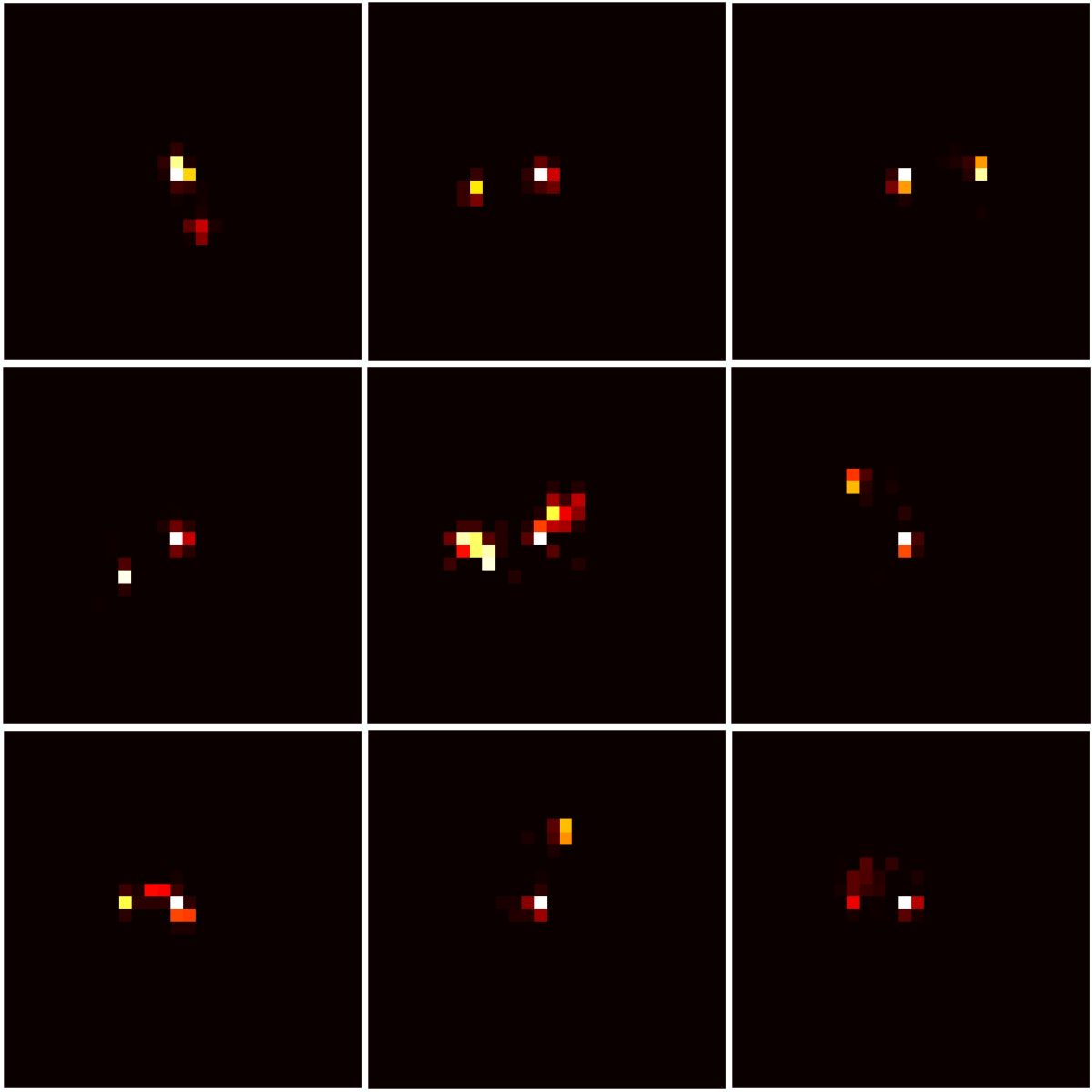}
\caption{}
\label{fig:gmsb_0_collage}
\end{subfigure}%
\begin{subfigure}{0.49\textwidth}
\centering
\includegraphics[width=0.95\textwidth]{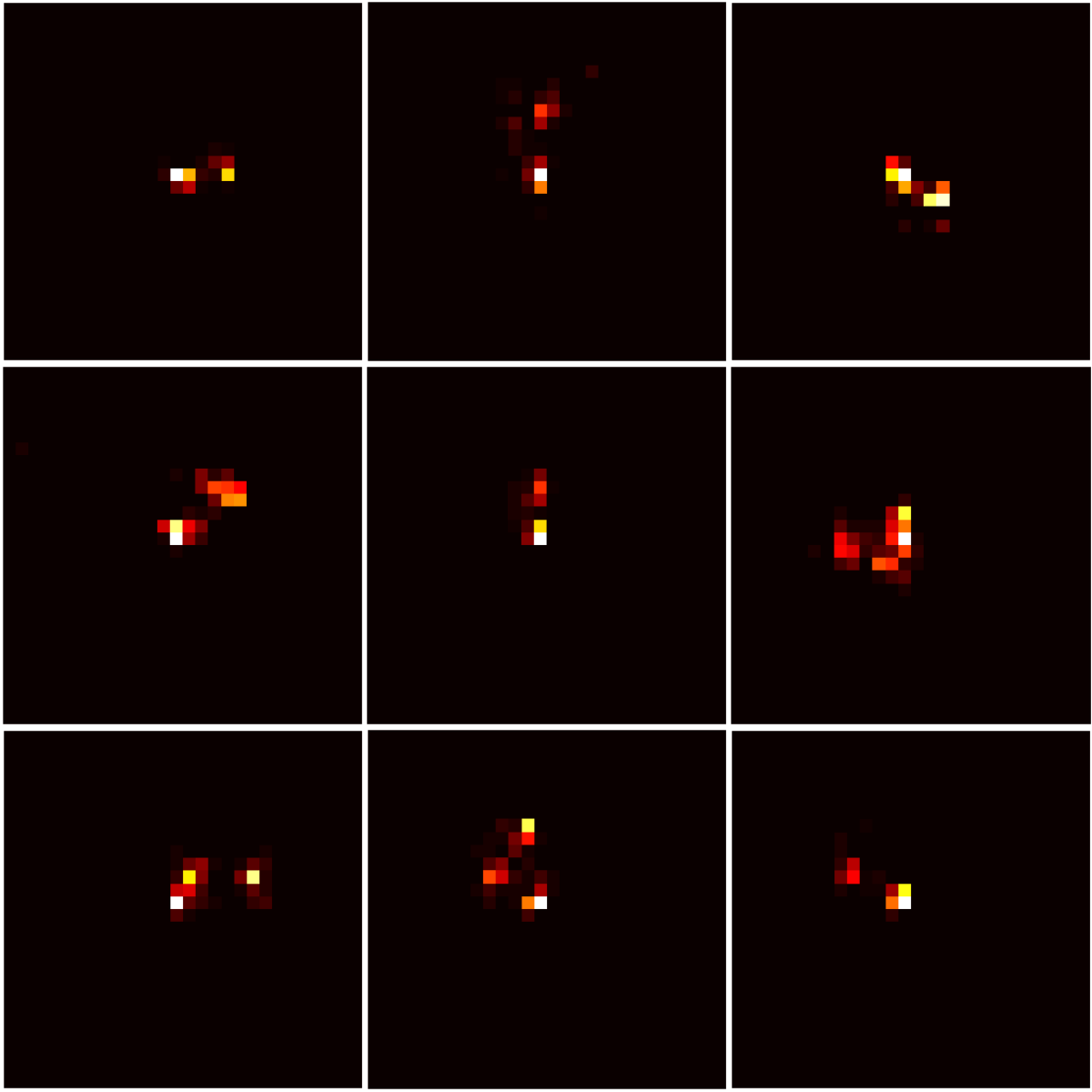}
\caption{}
\label{fig:gmsb_50_collage}
\end{subfigure} \\ 
\begin{subfigure}{0.49\textwidth}
\centering
\includegraphics[width=0.95\textwidth]{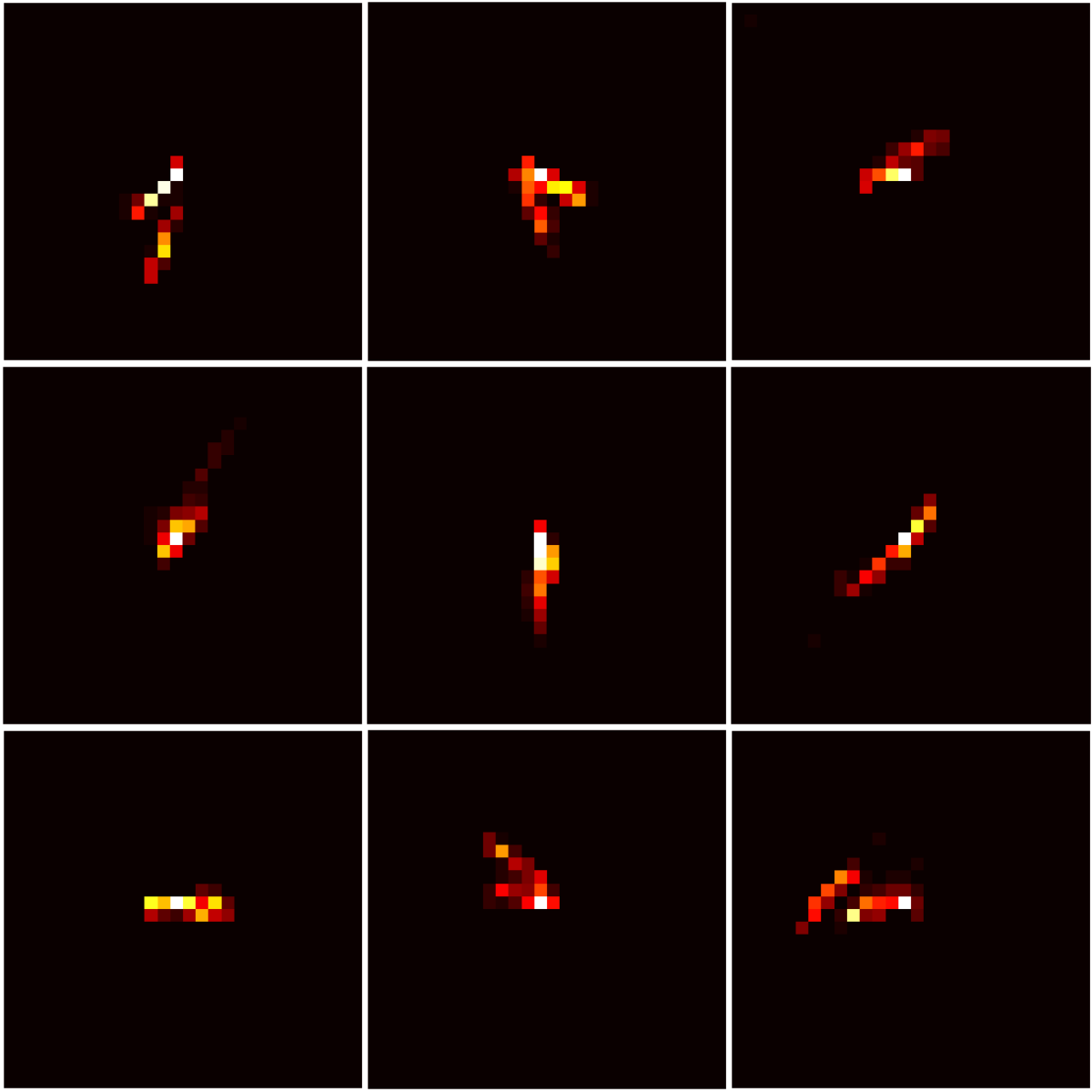}
\caption{}
\label{fig:gmsb_200_collage}
\end{subfigure}%
\caption{Typical energy deposition images of nine events from (a) non-displaced $Z$, (b) $Z$ having a transverse displacement between $50 {\rm~cm} - 70 {\rm~cm}$ and (c) $Z$ having a transverse displacement between $200 {\rm~cm} - 220 {\rm~cm}$. 
With increasing displacement, energy deposition becomes more and more elongated.}
\label{fig:gmsb_collage}
\end{figure}

Although statistically the difference between the displaced and prompt case is noticeable as is observed by the average images (Fig.~\ref{fig:avg_gmsb}), we need to check whether this difference is that prominent for individual images. Fig.~\ref{fig:gmsb_collage} shows nine typical images for jets coming from a prompt $Z$ (Fig.~\ref{fig:gmsb_0_collage}), and jets coming from $50-70 {\rm~cm}$ displaced $Z$ (Fig.~\ref{fig:gmsb_50_collage}) and from $200-220 {\rm~cm}$ displaced $Z$ (Fig.~\ref{fig:gmsb_200_collage}). We find that with increasing displacement, the energy deposits become more elongated and overall the multijet system is more contained. These features can easily be seen for the $200-220 {\rm~cm}$ displaced $Z$, but are not that prominent for the $50-70 {\rm~cm}$ displaced $Z$. Also, even the prompt case has some events where the energy depositions have some elongations. Therefore, it is not always possible to discriminate between them when we are given these individual images.

Standard cut-based analyses, therefore, might not be effective for this study. 
Even in the distributions shown in Fig.~\ref{fig:energy_gmsb}, the difference is present but subtle (note that the plots are shown in log scale). This motivates us to look for other powerful alternatives that can identify these subtle features and discriminate on the basis of that. We use image recognition technique by employing a CNN to check if this task can be achieved.
Before moving to the analysis using CNN, we study the difference in energy deposition patterns for the second scenario where the jets come directly from decay of LLP $X$.

\subsubsection{Displaced jets directly from LLP decay}
\label{sssec:disp_chi}

For the second 
scenario, we consider the decay of a boosted LLP $X$ 
into three quarks, giving multiple displaced jets coming from the LLP.
$$X\rightarrow j j j$$
To simulate this scenario we use R-parity violating (RPV) SUSY~\cite{Barbier:2004ez} where the neutralino is again the LLP and it can decay to three quarks via the $\lambda''_{ijk}$ (UDD) coupling.

We use \texttt{PYTHIA6} for generation of events and consider a non-displaced scenario along with the four cases with different displacements (listed in section \ref{list:disp_cases}). Also, to ensure that the decay products fall mostly within the calorimeter segment, we put a condition that the longitudinal displacement is less than $200 {\rm~cm}$ and $X$ is produced centrally ($|\eta|<0.7$). The mass of $X$ is taken to be $100 {\rm~GeV}$ (close to SM $Z$ boson mass to ensure that the multijet system here has similar boost as in the first scenario so that we can compare the two scenarios). We again take events where the HCAL energy deposit in $28\times28$ $i\eta-i\phi$ plane is between $(400,500){\rm~GeV}$ to make $X$ boosted enough such that its decay products come closer in the $\eta-\phi$ plane.

\begin{figure}[hbt!]
\centering
\begin{subfigure}{0.5\textwidth}
\centering
\includegraphics[width=7cm]{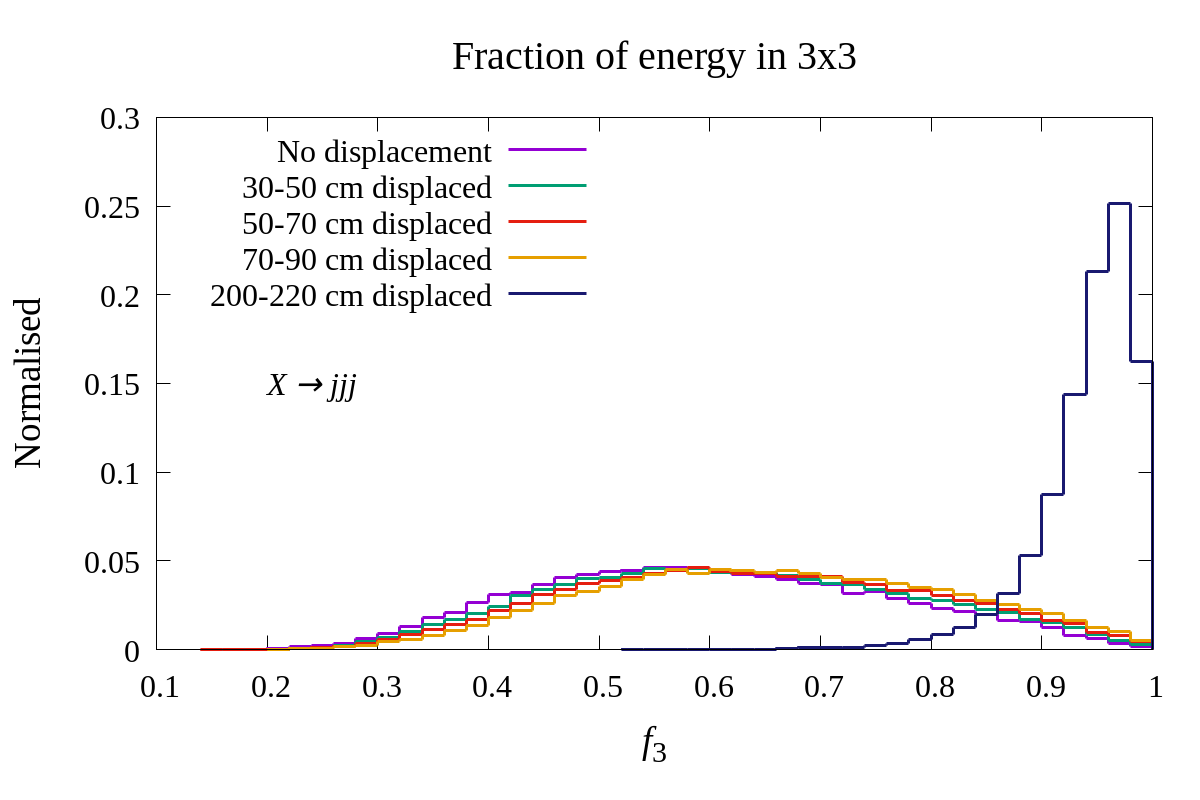}
\caption{}
\label{fig:3x3_udd}
\end{subfigure}%
\begin{subfigure}{0.5\textwidth}
\centering
\includegraphics[width=7cm]{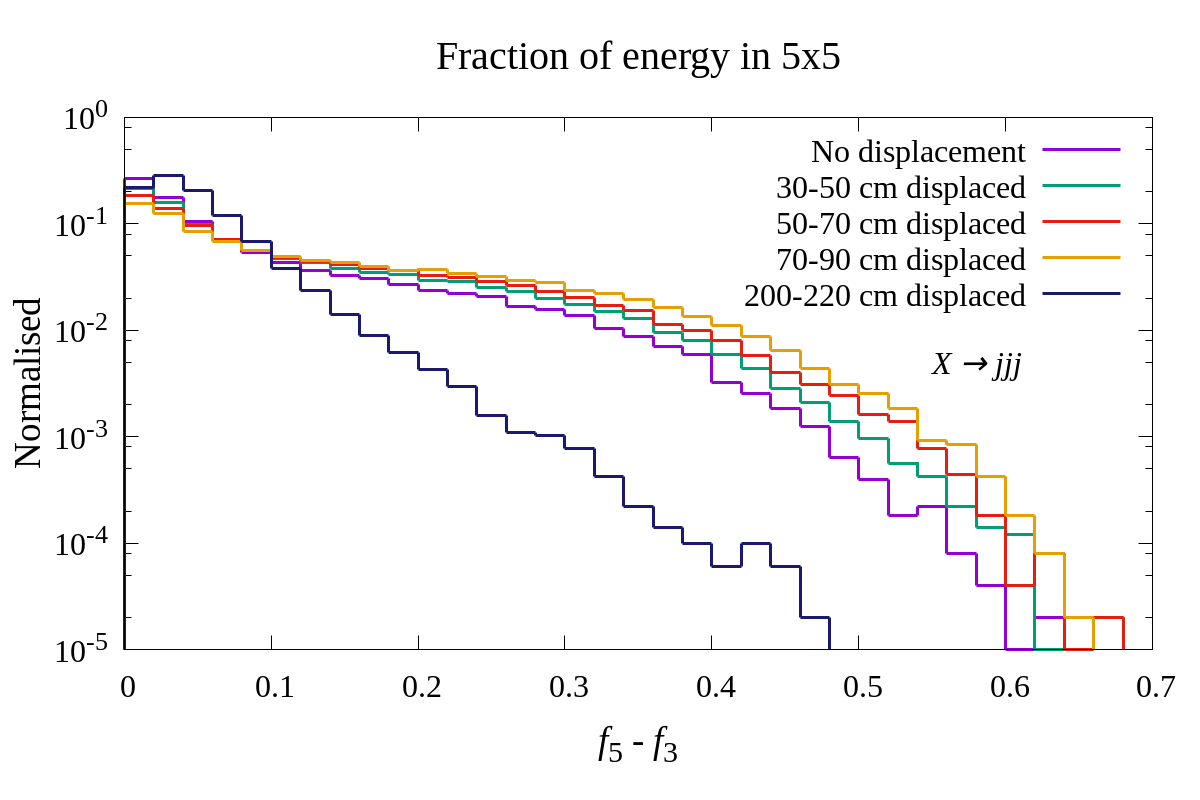}
\caption{}
\label{fig:5x5_udd}
\end{subfigure}\\
\begin{subfigure}{0.5\textwidth}
\centering
\includegraphics[width=7cm]{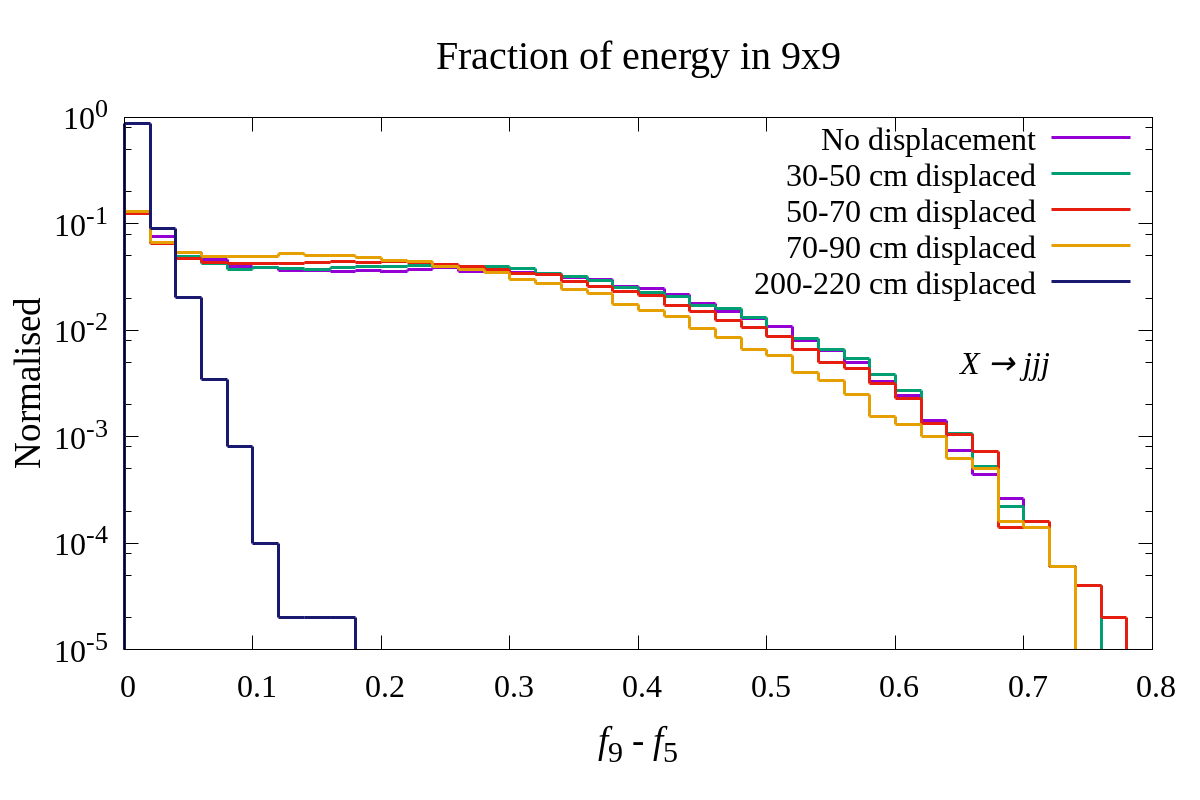}
\caption{}
\label{fig:9x9_udd}
\end{subfigure}%
\begin{subfigure}{0.5\textwidth}
\centering
\includegraphics[width=7cm]{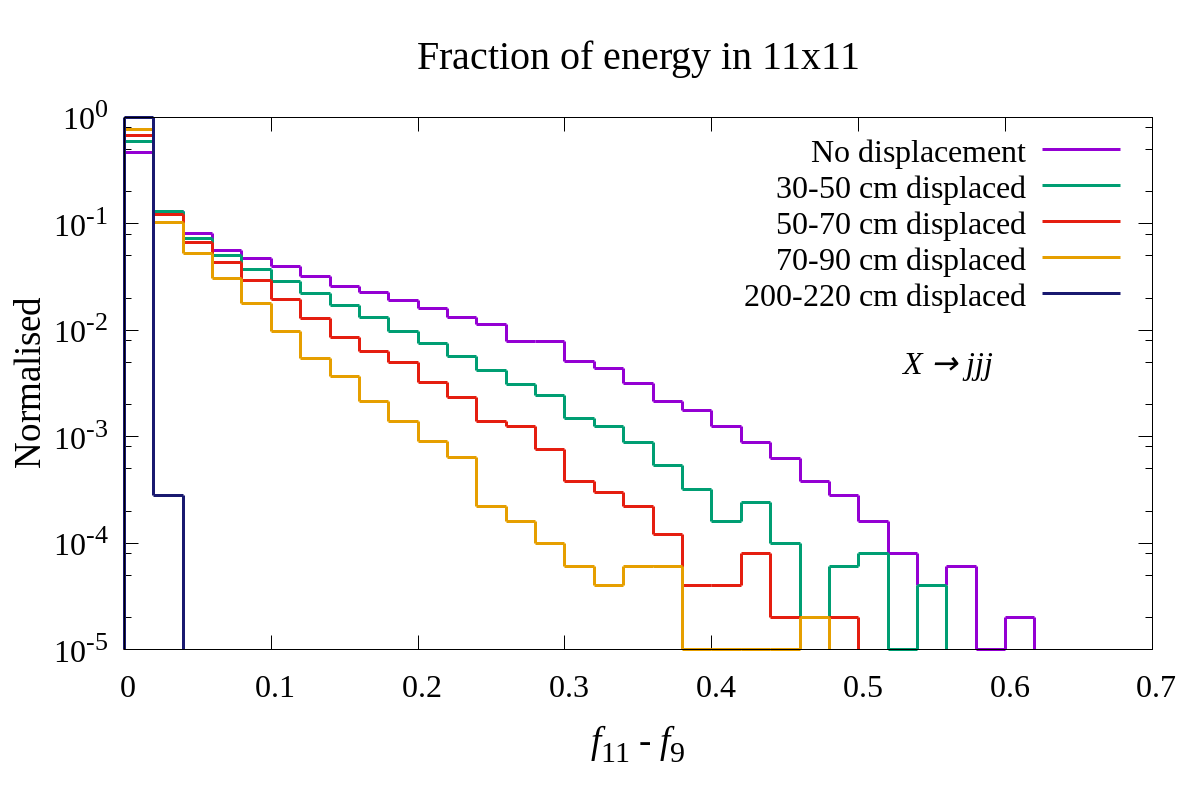}
\caption{}
\label{fig:11x11_udd}
\end{subfigure}%
\caption{Normalised distributions of energy deposition fraction ($f_i$) with varying sizes of blocks ($i\times i$). (a) Fraction of energy deposited in $3\times3$; (b),(c),(d) Fraction of energy deposited in $5\times5$, $9\times9$ and  $11\times11$ excluding the previous $3\times3$, $5\times5$ and $9\times9$ blocks' energy deposition fraction respectively.}
\label{fig:energy_udd}
\end{figure}

Fig.~\ref{fig:energy_udd} shows the normalised distributions of energy fractions for different scenarios discussed above. We observe that the cases where the decay happens within the Tracker have some slight differences with increasing the displacement gradually as we have seen in the previous scenario (section \ref{sssec:disp_Z}). And the $200 {\rm~cm} - 220 {\rm~cm}$ displaced case is very different from the other cases. There most of the energy is contained within the central $3\times3$ part of the $i\eta-i\phi$ plane.

\begin{figure}[hbt!]
\centering
\begin{subfigure}{0.5\textwidth}
\centering
\includegraphics[width=7.5cm]{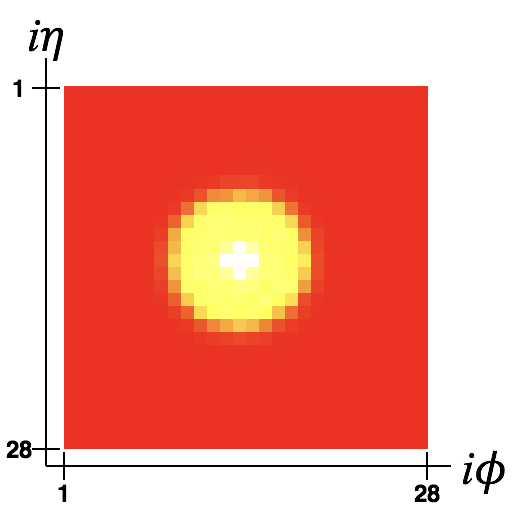}
\caption{}
\label{fig:0disp_udd}
\end{subfigure}%
\begin{subfigure}{0.5\textwidth}
\centering
\includegraphics[width=7.5cm]{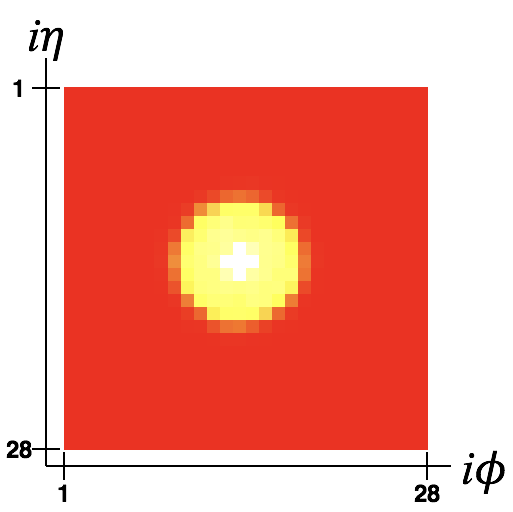}
\caption{}
\label{fig:30-50_udd}
\end{subfigure}\\
\begin{subfigure}{0.5\textwidth}
\centering
\includegraphics[width=7.5cm]{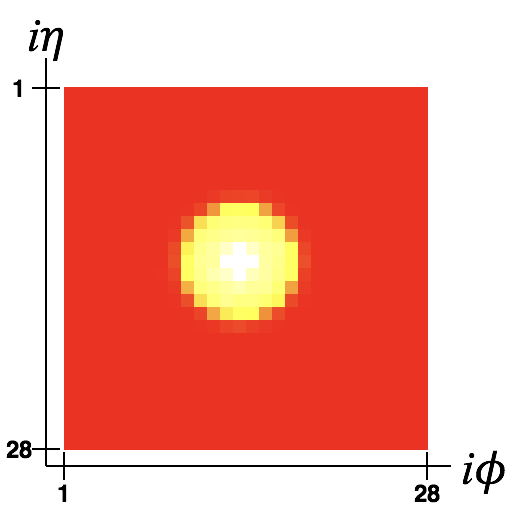}
\caption{}
\label{fig:50-70_udd}
\end{subfigure}%
\begin{subfigure}{0.5\textwidth}
\centering
\includegraphics[width=7.5cm]{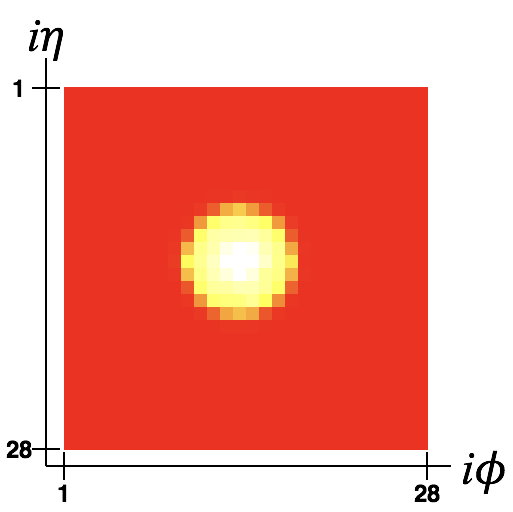}
\caption{}
\label{fig:70-90_udd}
\end{subfigure}%
\caption{Average of 50,000 images for (a) prompt jets from $X$; (b),(c),(d) jets displaced within $30-50 {\rm~cm}$, $50-70 {\rm~cm}$ and $70-90 {\rm~cm}$ transverse distance from PV respectively. These are $28\times28$ images in the $i\eta-i\phi$ plane with the highest energy bin in the center (14,14). With increasing displacement, the energy deposition is contained to smaller $i\eta-i\phi$ region. Here we do not show the 200-220 cm displaced case, for which the energy deposition is contained in a very small region.}
\label{fig:avg_udd}
\end{figure}

Fig.~\ref{fig:avg_udd} shows the average of 50,000 $X \rightarrow jjj$ images for each of the four cases of decay within the tracker. Here, since we have not applied any preprocessing,
 we cannot make out the three separate jets in the average images. We observe that with displacement the energy deposit becomes more contained in smaller $\eta-\phi$ region. As observed from the distributions in Fig.~\ref{fig:energy_udd}, for the maximum displaced case of $200-220 {\rm~cm}$, most of the energy deposition is contained within the first $3\times3$ block. The segmentation in the $z$ direction corresponding to $0.1$ $\eta$ segmentation gets bigger in physical size as one moves away from the IP radially (the segment size in $z$ direction for $\Delta\eta=0.1$ radian increases from A to C layers as can be seen in Fig.~\ref{fig:segment}). For the $200-220 {\rm~cm}$ displaced case, the decay happens just before entering the HCAL and therefore, the physical area taken by the decay products is very small and they mostly get contained within fewer $\eta-\phi$ towers. This is another important feature that we observe for displaced energy depositions.

This effect was also present in the displaced $Z$ case. It's more prominent for this case because here the $X$ is centrally produced and jets are coming directly from the decay of a boosted $X$ and are therefore, more collimated. So here, there is not much mismatch between the displaced particles' $\eta-\phi$ and the standard calorimeter segments. Therefore, 
the elongation of each objects'
energy deposition is very little and the energy deposition of the multijet system is more contained. For the previous case, the displaced jets come from the displaced $Z$ and even if the LLP $X$ is centrally produced, $Z$'s $\eta$ can have many possible values giving us more chances of mismatch with standard calorimeter segments, and hence the elongation feature is more. 


\clearpage

To summarise the last two sections (\ref{sssec:disp_Z} and \ref{sssec:disp_chi}), we identify two important features associated with displaced multijet systems. They are $-$
\begin{list}{$\bullet$}{}
\item \textbf{Elongated energy deposits in the HCAL} 

This happens due to the mismatch of displaced particles' $\eta-\phi$ direction with standard calorimeter $\eta-\phi$ towers. Therefore, energy deposition of displaced jets in the HCAL have more elongated patterns which differ from standard patterns of prompt jets.

\item \textbf{Total energy deposit of the multijet system more contained in the $i\eta-i\phi$ region}

The jets coming from the decay of the LLP have some $\Delta R$ \footnote{$\Delta R = \sqrt{\Delta\eta^2+\Delta\phi^2}$} between them. If the jets from $X$ have the same $\Delta R$ in both prompt decay as well as late decay of $X$, the energy deposit is smaller for the displaced case because the physical segmentation of the detector (in $z$ direction) has increased with increasing radial distance. 

\end{list}

The above features give different energy deposition patterns at the HCAL, but as seen from \ref{fig:energy_udd}, apart from the the 200-220 cm extremely displaced scenario, the difference between displaced and non-displaced cases are not significant enough 
so that we can use the usual cut-based analysis to discriminate them. On a qualitative level, it can also be seen from Fig.~\ref{fig:gmsb_200_collage}, where (a) and (b) categories are not distinguishable by human eye.
We therefore use the HCAL energy deposition images to train a convolutional neural network to learn these displaced features and discriminate displaced cases from prompt cases on the basis of that for both the LLP scenarios.


\subsection{The Convolution Neural Network}
\label{ssec:cnn}

The Convolutional Neural Network (CNN)~\cite{nonHEPref} is one of the most notable deep learning approaches used in diverse
computer vision applications. We use \texttt{Tensorflow}\cite{tensorflow} for implementing the CNN used in this work for discriminating displaced objects from non-displaced standard objects in the collider. As described in section \ref{ssec:segment_calo}, we have $28\times28$ 2-dimensional images of HCAL energy deposition for all scenarios. 
These images are the input of our neural network. In the next subsection we briefly discuss the various components of the network. 

\begin{figure}[hbt!]
\centering
\includegraphics[width=\textwidth]{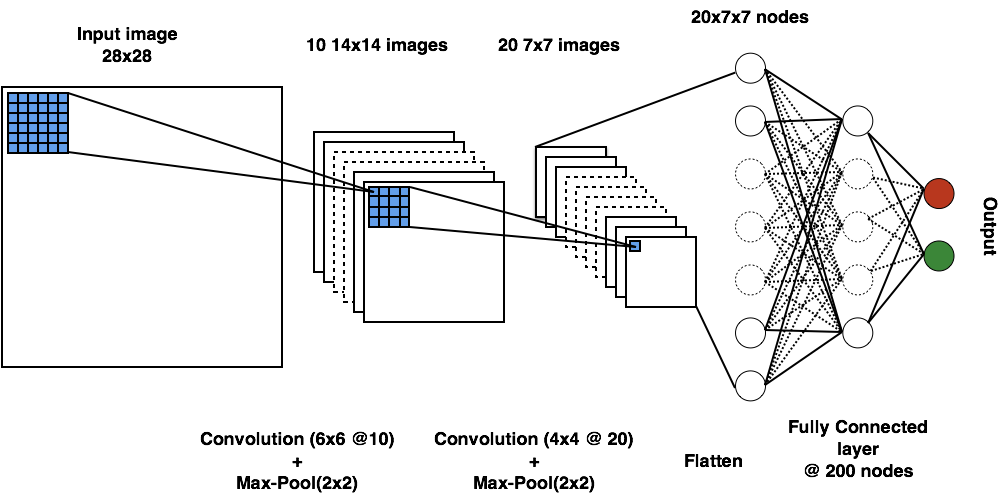}
\caption{The CNN architecture used.}
\label{fig:nn_architect}
\end{figure}

\subsubsection{Network Architecture}
\label{sssec:net_arch}

The architecture used in our work is inspired from some of the previous works in high energy physics where they tag hadronic decays of boosted objects like $W$ boson and top quark using CNN \cite{deOliveira:2015xxd,Kasieczka:2017nvn}, because we are also studying boosted jet systems. However, we have tuned the hyperparameters for better performance in our case.

\begin{enumerate}
\item \textbf{Convolution Layer:} In convolutional layers, the algorithm utilises various kernels to convolve the whole
image to generate various feature maps. We use two convolution layers: 
\underline{Layer 1:} 10 filters of kernel size $6\times6$; and  \underline{Layer 2:} 20 filters of kernel size $4\times4$, with a stride of $1\times1$ for both the layers, which means 
that the filter convolves around the input volume by shifting one unit at a time. 
The objective of the convolution operation is to extract features such as edges and shapes. 
In order to introduce nonlinearity to the system, activation function of Rectified Linear Unit (RELU)\cite{Nair:2010:RLU:3104322.3104425} has been applied to the outputs; and L2 regularization~\cite{AndrewNg} has been applied to the kernel weights. The outputs of the convolution layer are also batch normalised\cite{Ioffe2015BatchNA}.

\item \textbf{Max-Pool:} After each convolution layer, the output has been max-pooled with a pool size of $2\times2$ which means that each $2\times2$ kernel of the convolution output has been replaced by the maximum value in that kernel. This reduces the dimension of the image by half after each max-pooling and finally we have 20 
$7\times7$ images after convolution and max-pooling. By now, we have enabled the model to understand the features. Next, we will flatten the output and feed it to a fully-connected Neural Network for classification purposes.

\item \textbf{Flatten and Fully Connected (FC) Layer:} At this stage, we flatten these images and get $20\times7\times7$, i.e., 980 input values which we connect to a fully connected layer with 200 nodes. Activation function RELU is applied. We apply a $50\%$ dropout\cite{JMLR:v15:srivastava14a} to this FC layer to deal with the problem of overfitting. Finally this layer is connected to the binary output through a softmax activation function~\cite{softmax}.
\end{enumerate}

Fig.~\ref{fig:nn_architect}~summarises the network architecture used in this work. We use Adam Optimizer\cite{Kingma2015AdamAM} with a learning rate of 0.001. We train to minimise the cross-entropy loss function~\cite{xEntropy}. We have done some naive optimisations to decide on the present network architecture and hyperparameters.

For simplification in our work, the energy deposition in 4 radial HCAL layers are added together to get the total energy deposition. In this way we are able to use 2D filters on 2D images. However, for better separation power, one can use 3D image by taking into account energy deposits in each radial layer separately or considering 4 different channels for each pixel with the energy deposited values in each layer as one channel. Further discussion on energy deposition in each layer can be found in appendix \ref{app:layers}. In the former case, 3D filters will be needed, and the training process would be more resource consuming for both the cases.
We also limit ourselves to images of HCAL energy deposits in $\eta$-$\phi$ plane and feed those images to a CNN, while more complicated setups can be used for an actual search by experimental collaborations, in order to achieve better discrimination power between signal and background. For example, a hybrid CNN model can be built, that takes additional input variables besides the image in the first fully connected layer after flattening. 
These additional input variables can come from other parts of the detector like tracker and ECAL. These variables could be number of tracks below a calorimeter energy deposit, energy deposited in the ECAL or the ratio of ECAL and HCAL energy deposits, and these can increase the sensitivity to identify displaced objects. Also we can use separate CNNs with different architectures on different parts of the detector (like separate CNNs for ECAL and HCAL energy deposition images) and add them after flattening because understanding features from different detector parts might require different levels of complexity in terms of the network architecture \cite{Lin:2018cin,Kim:2019wns,Chen:2019uar}.

\subsubsection{Dataset}
\label{sssec:dataset}

We have used 100,000 images (including both classes) out of which 60,000 images have been used for training, 20,000 for validation and another 20,000 for testing the network. We use a batch size of 200 while training. The training was stopped at the epoch with minimum validation loss when this loss was not decreasing significantly.

We performed the classification between the non-displaced case and the four different displaced cases for each of the two scenarios $-$ displaced $Z$ from $X$ which decays to jets and displaced jets coming from $X$ directly.


\subsection{Analysis and Results}
\label{ssec:analysis}

For analysing the performance of the network, we draw a Receiver Operating Characteristics (ROC)\cite{FAWCETT2006861} curve with signal efficiency vs. background rejection. Since we want to study the difference in energy deposition patterns as a result of displacement of particles, here, we consider the non-displaced events as our background and the four displaced cases as separate signals. We use the test output of the CNN to draw the ROC curves. These curves will give us an idea of the discriminating power of the classifier for different displacements and for which signal it works the best.

\subsubsection{Displaced jets from displaced $Z$}
\label{sssec:Z_ROC}

\begin{figure}[hbt!]
\centering
\includegraphics[width=12cm]{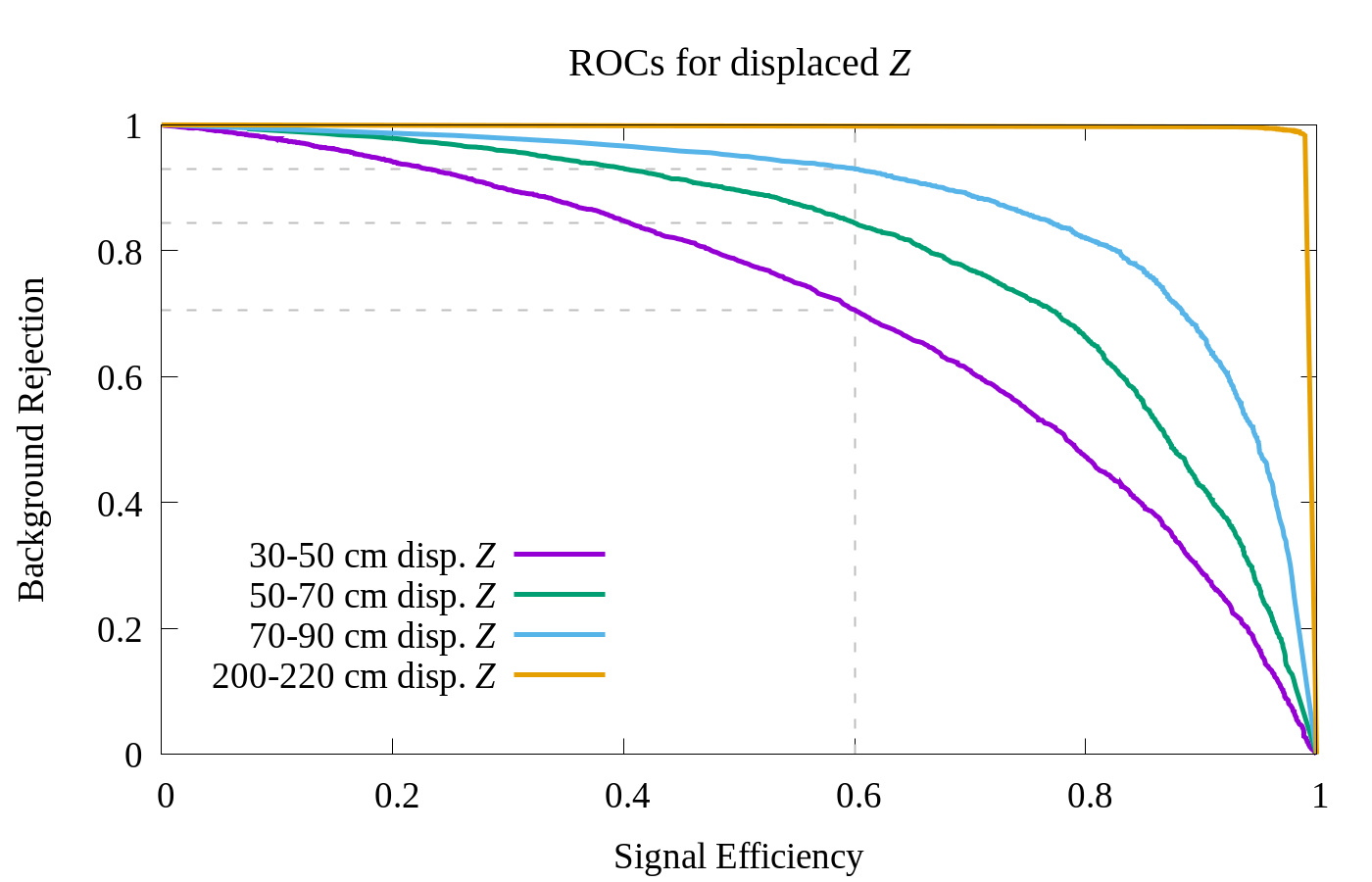}
\caption{ROCs of the CNN performance to separate non-displaced $Z$ from different classes of displaced $Z$.}
\label{fig:ROC_gmsb}
\end{figure}

Fig.~\ref{fig:ROC_gmsb} shows the ROCs of the CNN performance for the non-displaced vs the four cases with different displacements of $Z$ boson.

We get background rejection of $70.61\%$, $84.36\%$ and $93.02\%$ for a signal efficiency of $60\%$ for $30 {\rm~cm} - 50 {\rm~cm}$ displaced $Z$, $50 {\rm~cm} - 70 {\rm~cm}$ displaced $Z$, and $70 {\rm~cm} - 90 {\rm~cm}$ displaced $Z$ respectively. We notice that the performance of the network is better for more displaced $Z$ which is expected. 
For the $200 {\rm~cm} - 220 {\rm~cm}$ displaced $Z$, the network performs the best, and we get background rejection of $99.8\%$ for a signal efficiency of $60\%$. This implies that the network has learned the features associated with displacement and discriminates on the basis of that. For the most displaced case, there is more mismatch between the decay products $\eta-\phi$ with the standard $\eta-\phi$ HCAL towers, hence more elongated energy deposition in the HCAL. We therefore find that this analysis is better for more displaced scenarios, where usually our standard reconstructions fail badly.

\begin{figure}[hbt!]
\centering
\includegraphics[width=12cm]{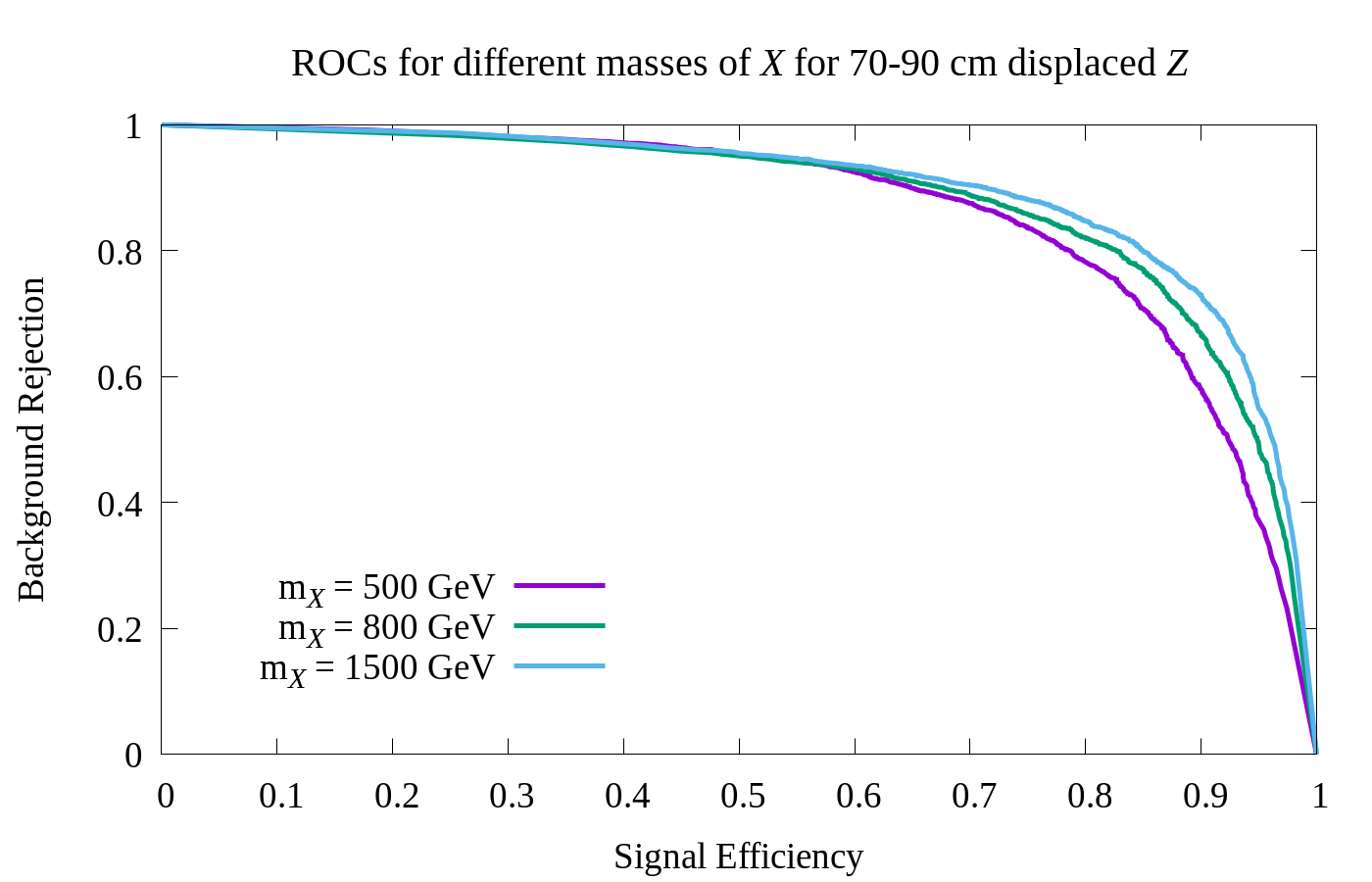}
\caption{ROCs of the CNN performance to separate non-displaced $Z$ from $70{\rm~cm}-90{\rm~cm}$ displaced $Z$ for different masses of $X$.}
\label{fig:ROC_gmsb_mass}
\end{figure}

In the above analysis the mass of $X$ was taken to be $800 {\rm~GeV}$. We now study the dependence of this analysis on mass of the LLP. We consider the performance of the network to discriminate $70-90 {\rm~cm}$ displaced $Z$ from prompt $Z$ for three different masses of $X$ $-$ $500 {\rm~GeV}$, $800 {\rm~GeV}$ and $1500 {\rm~GeV}$. We choose the same energy window cut $(400,500) {\rm~GeV}$ for all these cases. Hence, the boost of the multijet
system coming from $Z$ remains the same for all the cases. Fig.~\ref{fig:ROC_gmsb_mass} shows the ROCs for the CNN performance to separate non-displaced $Z$ from $70{\rm~cm}-90{\rm~cm}$ displaced $Z$ for these different $X$ masses. We find that with increasing $X$ mass the CNN performs better, although the effect is not drastic.


Massive LLPs travel slower in the detector and therefore their decay products can have large $\Delta R$ between them. Large $\Delta R$ between the $Z$ and the invisible particle means that the jets from $Z$ pass through different standard $\eta-\phi$ calorimeter towers in different layers and hence the projection along constant $\eta-\phi$ of these radial layers will have more elongated energy deposition pattern. This improves the discrimination power of the CNN with increasing LLP mass.



\subsubsection{Displaced jets directly from LLP decay}
\label{sssec:chi10_ROC}

\begin{figure}[hbt!]
\centering
\includegraphics[width=12cm]{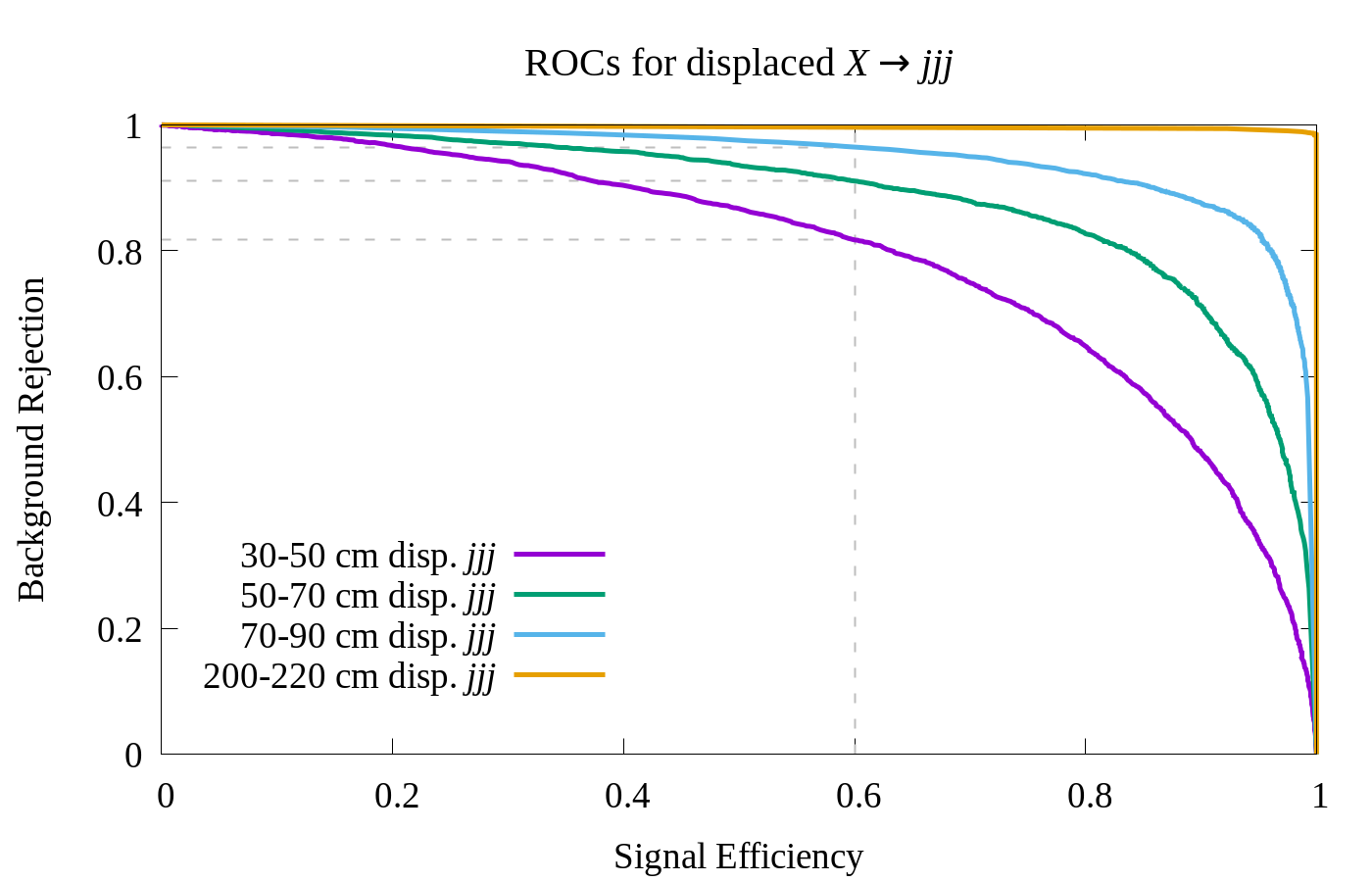}
\caption{ROCs of the CNN performance to separate non-displaced $jjj$ from different classes of displaced $jjj$.}
\label{fig:ROC_udd}
\end{figure}

Fig.~\ref{fig:ROC_udd} shows the ROCs of the CNN performance for the non-displaced vs the four cases with different displacements of the jets coming from the decay of $X$.

We get background rejection of $81.70\%$, $91.06\%$ and $96.39\%$ for a signal efficiency of $60\%$ for $30 {\rm~cm} - 50 {\rm~cm}$ displaced, $50 {\rm~cm} - 70 {\rm~cm}$ displaced, and $70 {\rm~cm} - 90 {\rm~cm}$ displaced decay of $X$ respectively. Again the performance of the network is better for more displaced cases, being the best for the $200 {\rm~cm} - 220 {\rm~cm}$ displaced decay of $X$, and we get background rejection of $99.6\%$ for a signal efficiency of $60\%$.

Varying the mass of the LLP in this scenario needs different energy deposition window cut (other than $(400,500) {\rm~GeV}$) for giving the same amount of boost to the final multijet system. We believe that if the multijet system  has the same boost, the results won't be affected with variation of LLP mass for this scenario.

Therefore, we find that CNNs can learn displaced features from HCAL energy 
deposition images and is able to discriminate prompt multijet systems from displaced ones. We have repeated this analysis by also considering images of transverse energy deposition in the HCAL. Our results are robust against this change.

The two scenarios considered by us had some kinematic differences. Yet the network performs equally well for both the scenarios with increasing displacement. This suggests that this kind of analysis is quite robust to the LLP model that we consider and hence can be extended to study other LLP decaying to multijet scenarios as well.


\begin{figure}[hbt!]
\centering
\begin{subfigure}{0.49\textwidth}
\includegraphics[height=4.5cm]{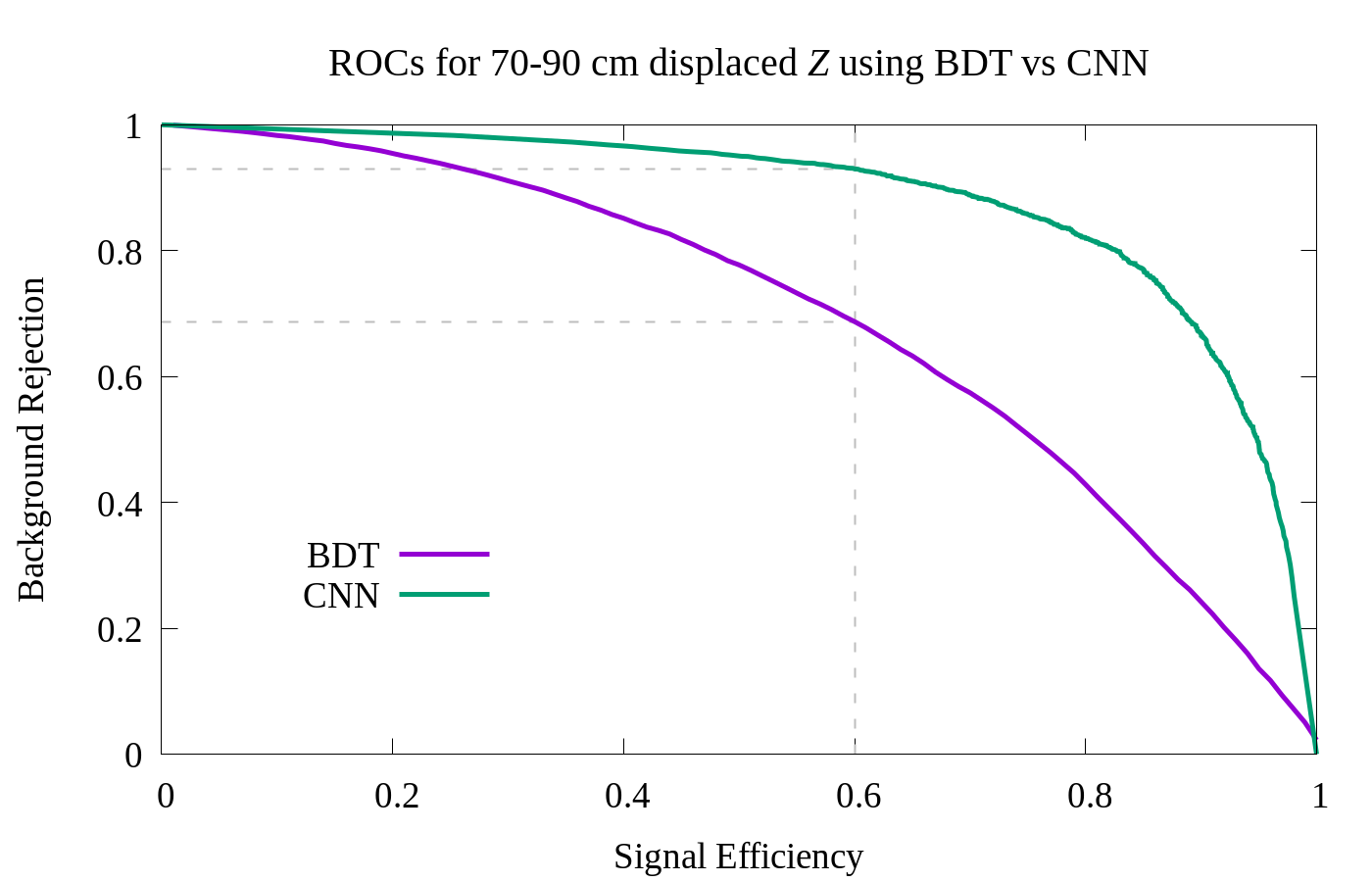}
\label{fig:gmsb_BDT}
\end{subfigure}%
\begin{subfigure}{0.49\textwidth}
\centering
\includegraphics[height=4.5cm]{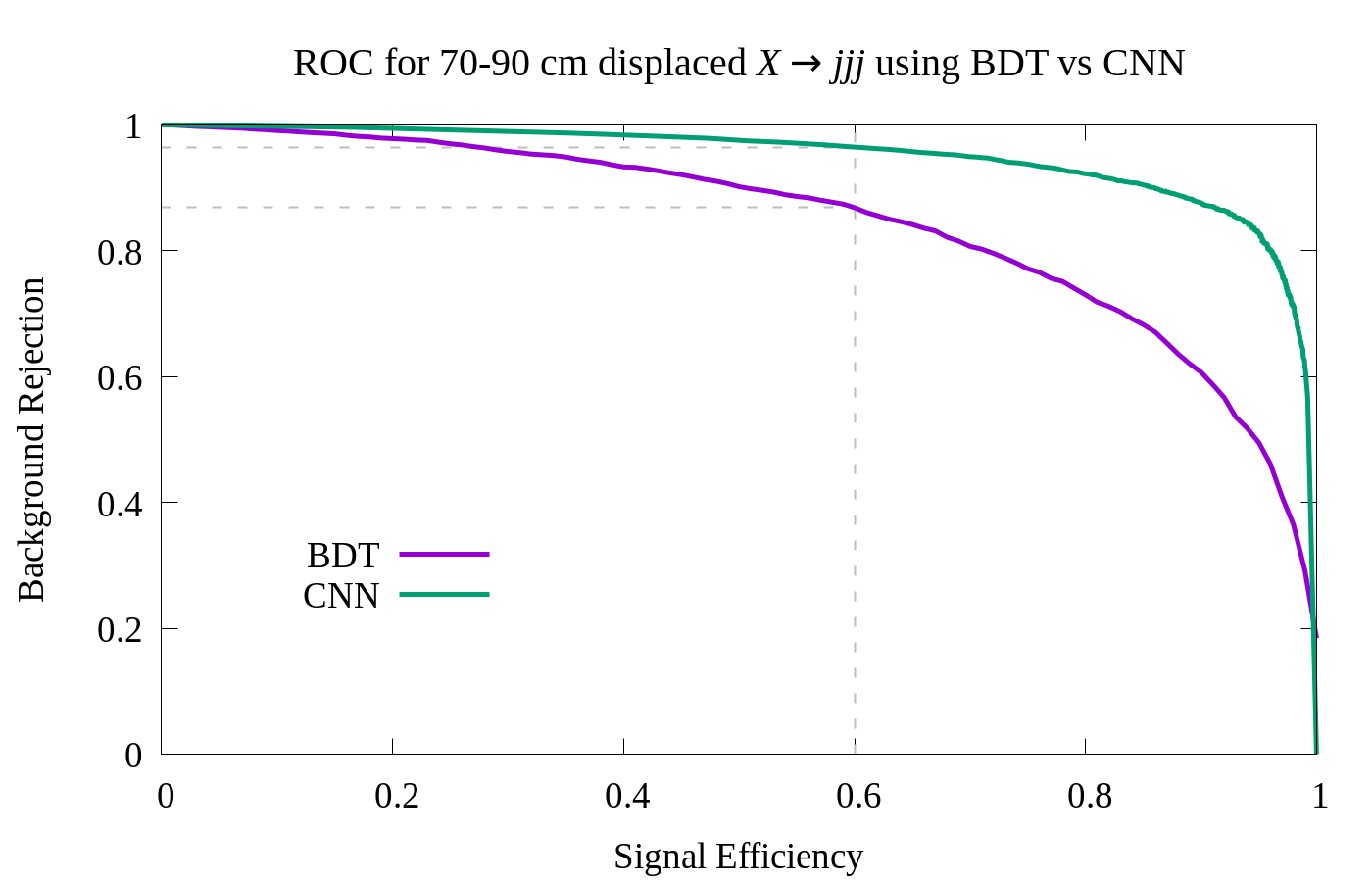}
\label{fig:udd_BDT}
\end{subfigure}
\caption{
CNN vs BDT: comparison of performance for both models.
}
\label{fig:BDT_compare}
\end{figure}

As a benchmark to which one can compare the performance of CNN, 
we have performed a similar classification exercise using boosted decision tree (BDT) with variable decorrelation. The energy fraction variables $f_3$, $f_5$, $f_9$ and $f_{11}$ were used as input to the framework. 
For both models, (1) jets coming from a displaced Z boson and (2) jets coming directly from the decay of an LLP, we found CNN to work much better than standard BDT based on the energy fraction variables. The plots 
comparing the performance of BDT and CNN is presented in Fig.~\ref{fig:BDT_compare}. 
It was found that for $60\%$ signal efficiency, we obtain a $68.73\%$ background rejection using BDT, while this was $93.02\%$ using CNN for the first case. For the second case, we obtain a $86.84\%$ background rejection using BDT, and $96.39\%$ background rejection using CNN for a $60\%$ signal efficiency. It is also important to note that the performance of CNN is even better for higher signal efficiencies than BDT.

In the next section, we provide a brief discussion of a special case of $X \rightarrow jjj$ where $X$ decays at rest. This resembles the case of a color neutral R-hadron stopping in the detector and decaying into jets. We study whether HCAL images have any potential for stopped R-hadron studies.

\subsection{Stopped particle scenario}
\label{sssec:stopped}

Particles with longer lifetime can occur in split supersymmetry~\cite{Farrar:1996rg,Baer:1998pg,Mafi:1999dg,Kraan:2004tz,ArkaniHamed:2004fb,Giudice:2004tc,Hewett:2004nw,Arvanitaki:2005nq,Arvanitaki:2012ps}, where the decay of gluino ($\tilde{g}$) is suppressed due to the large mass difference between gluino and squark; i.e. squark is much heavier than gluino in this model. If long-lived gluinos exist they might be produced in the $pp$ collisions in the LHC, and they will soon hadronise to make a hadron-like state, generally referred to as ``R-hadrons''. These R-hadron can be charged or neutral, and they will lose energy by interacting with the material of the detector as they travel through it. For heavy R-hadrons, which will move slowly, the energy loss will be sufficient to stop a significant fraction of the produced R-hadrons inside the calorimeter of the detector. These ``stopped'' particles may decay seconds, minutes, hours, or days later, resulting in out-of-time energy deposits in the calorimeter. The latest R-hadron search results at CMS and ATLAS are shown in \cite{Sirunyan:2017sbs,Aaboud:2019trc} respectively. 


We want to explore how the HCAL energy deposition pattern would look like for a stopped particle scenario. We expect this to be quite different from standard deposition patterns. Since the stopped particle decays at rest at a significant distance from the PV, $\eta$ and $\phi$ of the decay products don't match with the HCAL $\eta-\phi$ segmentation. The energy deposit of these particles, therefore, won't be contained in one or two $\eta-\phi$ towers of the HCAL. We rather expect these energy deposits to look like lines on the $\eta-\phi$ plane.

To demonstrate this, we consider the decay of $X$ of mass $1 {\rm~TeV}$ to three quarks as described for the 
second scenario above. 
We use \texttt{PYTHIA6} and RPV SUSY model to simulate this case. But here the LLP $X$ is made to decay at rest. The position where $X$ stops is simulated such that it follows an exponential distribution. We consider events where $X$ has stopped in the first HCAL layer.
After $X$ decays, we will get some particles moving in the forward direction as well as some in the backward direction to conserve momentum. The backward moving particles have very unique signatures in the collider and is in itself a very interesting subject of study \cite{Banerjee:2017hmw}. In the present work, we only consider the energy deposition of forward-moving particles. Therefore, most of the events have energy deposition in the HCAL between $(400,500) {\rm~GeV}$, which is about half of the $X$ mass.

\begin{figure}[hbt!]
\centering
\begin{subfigure}{0.49\textwidth}
\includegraphics[height=7cm]{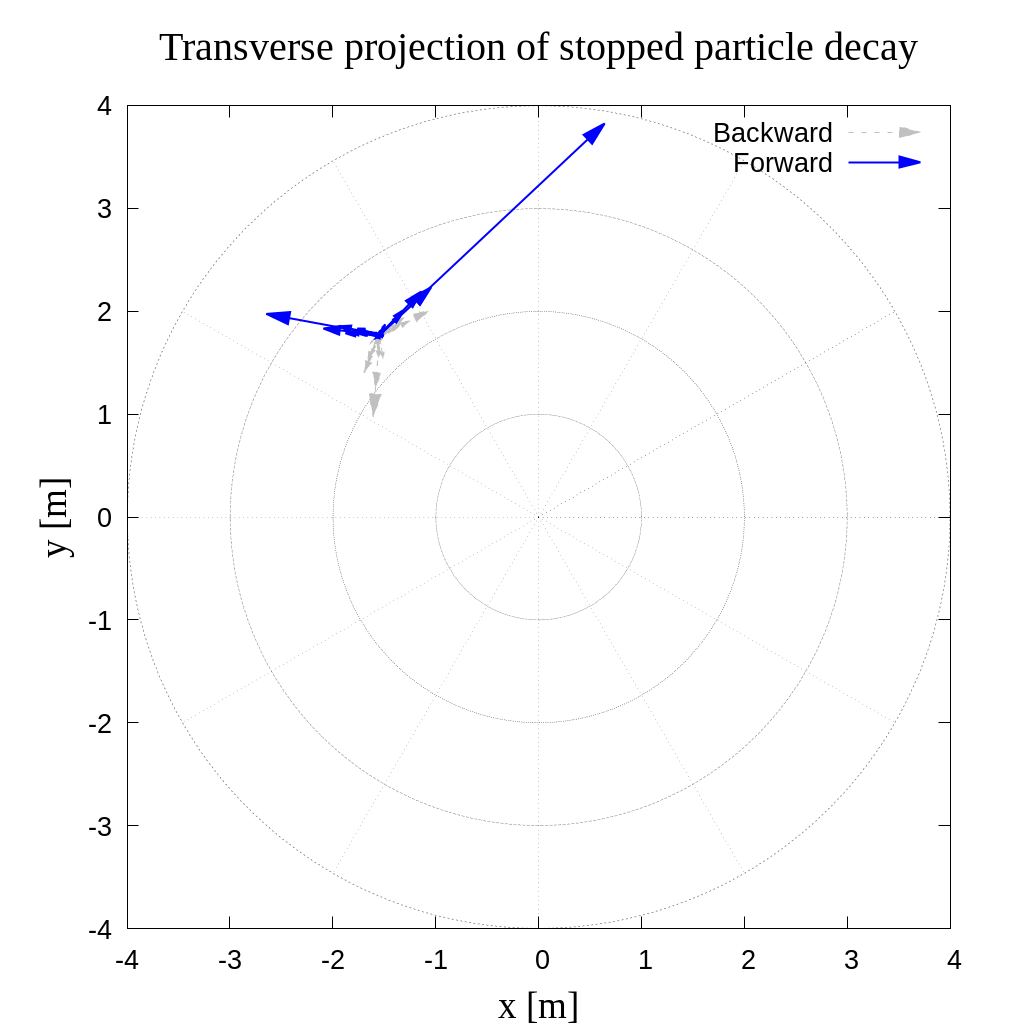}
\caption{}
\label{fig:stopped_xy_pro}
\end{subfigure}%
\begin{subfigure}{0.49\textwidth}
\centering
\includegraphics[height=7cm]{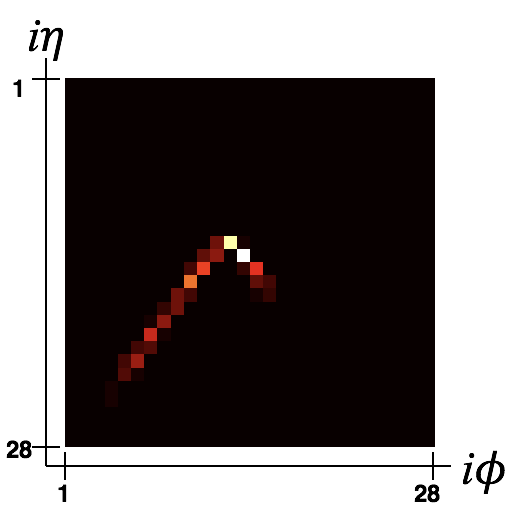}
\caption{}
\label{fig:stopped_HCAL}
\end{subfigure}
\caption{(a) Stopped particle decay projection in the transverse plane $-$ shown in grey and blue are particles moving backward and forward respectively; (b) Energy deposition of forward-moving particles coming from the decay of a stopped particle in the HCAL. Both images are for the same stopped particle decay event.}
\label{fig:stopped}
\end{figure}

Fig.~\ref{fig:stopped_xy_pro} shows the $x$-$y$ (transverse) projection of a typical stopped $X$ decay. Marked in blue are the particles moving in the forward direction which we have propagated through our segmented HCAL. Fig.~\ref{fig:stopped_HCAL} shows the energy deposit of forward-moving particles for the same event in $i\eta-i\phi$ plane. We find that this energy deposition pattern is quite different from standard scenarios. Inclusion of backward-moving particles will enhance this feature and we will get more such lines.

Since top quarks can decay to three quarks, we have performed a classification between stopped $X$ decaying to three jets and top quark using the CNN architect as described in section \ref{ssec:cnn} with energy of top lying between $(400,500) {\rm~GeV}$, same as the energy range for stopped case. They have very different energy deposition patterns in the HCAL and as expected we get high accuracy from the CNN training as well as validation. Even for $95\%$ stopped particle tagging efficiency, we get $\sim 99\%$ single top background rejection efficiency.




\section{Conclusion and Outlook}
\label{sec:concl}

This work presents an idea of how HCAL energy deposition images along with image recognition techniques can be used in the search for long-lived particles to distinguish between prompt and displaced jets. To the best of our knowledge, this work is the first  attempt in studying LLPs using energy deposition images and image recognition techniques. LLPs are difficult to identify using standard reconstructions due to their displacement from the PV. In this work we propose an additional method which can be used in combination with other standard LLP studies. We consider two scenarios which are different in the sense that in one, the displaced jets come from the decay of an intermediate displaced SM particle ($Z$ boson) while in the other they directly come from the decay of the LLP. By studying the energy deposition patterns of LLPs with varying displacements, in these two scenarios, we observe two key features. One is, elongation in energy deposition because $\eta$ and $\phi$ of particles which are very much displaced from the PV do not match with standard detector $\eta-\phi$ segmentation. Another one is, later the decay of the LLP, smaller is the physical region in which the energy deposition of its decay products is contained in HCAL. Due to the absence of layered structure and $z$ segmentation in fast detector simulations like \texttt{Delphes}, we can't use them to study these features for displaced jets. Therefore, we simulate our simplified calorimeter following the segmentation of the Tile Calorimeter of ATLAS.

We used these displaced features of LLPs that give different energy deposition patterns in the HCAL to differentiate them from non-displaced objects using a convolutional neural network. Our analysis performs better for LLPs which decay at  larger distances from the PV, where usual displaced jets analysis might lose sensitivity due to failure of standard reconstructions. Therefore, this might serve as a complementary analysis technique to standard LLP analyses techniques; or this method can be used in conjunction with other relevant information from tracker, ECAL and muon system.

As a limiting case of the second scenario where the jets come directly from the LLP decay, we consider the situation where the LLP stops before decaying. We show that stopped particles also have very different energy deposition patterns in the calorimeter. Therefore, we can consider to look for such HCAL images rather than waiting for empty bunch crossings for the search of stopped R-hadrons. 

We have used minimal preprocessing to the images and have not done advanced optimisations. We would like to reiterate here that the major focus of this study was to show the feasibility of probing displaced jets emanating from the decay of an LLP via ML techniques. Advanced pre-processing and optimisations can be done for dedicated LLP searches. In this study a simplified detector simulation is used, which only accounts for geometry. There is no realistic calorimeter simulation with lateral and longitudinal shower shapes to account for overlapping showers, and there is no pile-up included. These are beyond the scope of this study.

Although we have shown this image-based analysis technique for some particular scenarios in this work, we believe that it will work for any scenario where an LLP decays into multiple jets. Therefore, this can be treated as a robust search technique for LLPs decaying to give displaced jets in the final state.
This work is a simple-minded analysis done for the proof-of-principle that displaced jets have some different energy deposition features and these can be identified using an image based study.

Lastly, with rapid advancement in the field of deep learning, several other new methods have come up recently which could give similar or better results than CNN. Capsule Neural Network (CapsNet) \cite{Diefenbacher:2019ezd} aims to make improvements to CNN by handling the spatial relationship in an image more efficiently. On the other hand, PointNet$++$ is a pioneering work in applying machine learning on point clouds \cite{Komiske:2018cqr,Qu:2019gqs}, which is a collection of high dimensional objects. Another important ML method, that could be explored in the context of displaced objects, is Graph Neural Network (GNN), which directly operates on the graph structure. GNNs have found extensive use in many other high-energy physics applications \cite{Abdughani:2018wrw,Martinez:2018fwc,Ren:2019xhp}. \\

{\it Acknowledgements: }
The work of B.B. is supported by the Department of Science and Technology, Government of India, under Grant No. IFA13- PH-75 (INSPIRE Faculty Award). 
The work of S.M. is supported by the German Federal Ministry of Education and Research BMBF. 
R.S. would like to thank Rahool Kumar Barman and Amit Adhikary for useful discussions.


\appendix

\section{Validation of our HCAL segmentation with \texttt{Delphes}}
\label{app:val}

We present here a small validation of our HCAL segmentation with standard fast detector simulation $-$ \texttt{Delphes}. To the best of our knowledge there are no ATLAS public results using which we can cross check our results for long-lived particles. Since \texttt{Delphes} has been parametrized with ATLAS for prompt particles \cite[section 5]{deFavereau:2013fsa}, we can validate our segmentation with \texttt{Delphes} using prompt jets. As we have discussed earlier that \texttt{Delphes} will give the $\eta$-$\phi$ of the particles coming from the LLP decay with respect to the secondary vertex, rather than the $\eta$-$\phi$ in which they will actually be detected according to the detector segmentation.

\begin{figure}[hbt!]
\centering
\begin{subfigure}{0.5\textwidth}
\centering
\includegraphics[width=8cm]{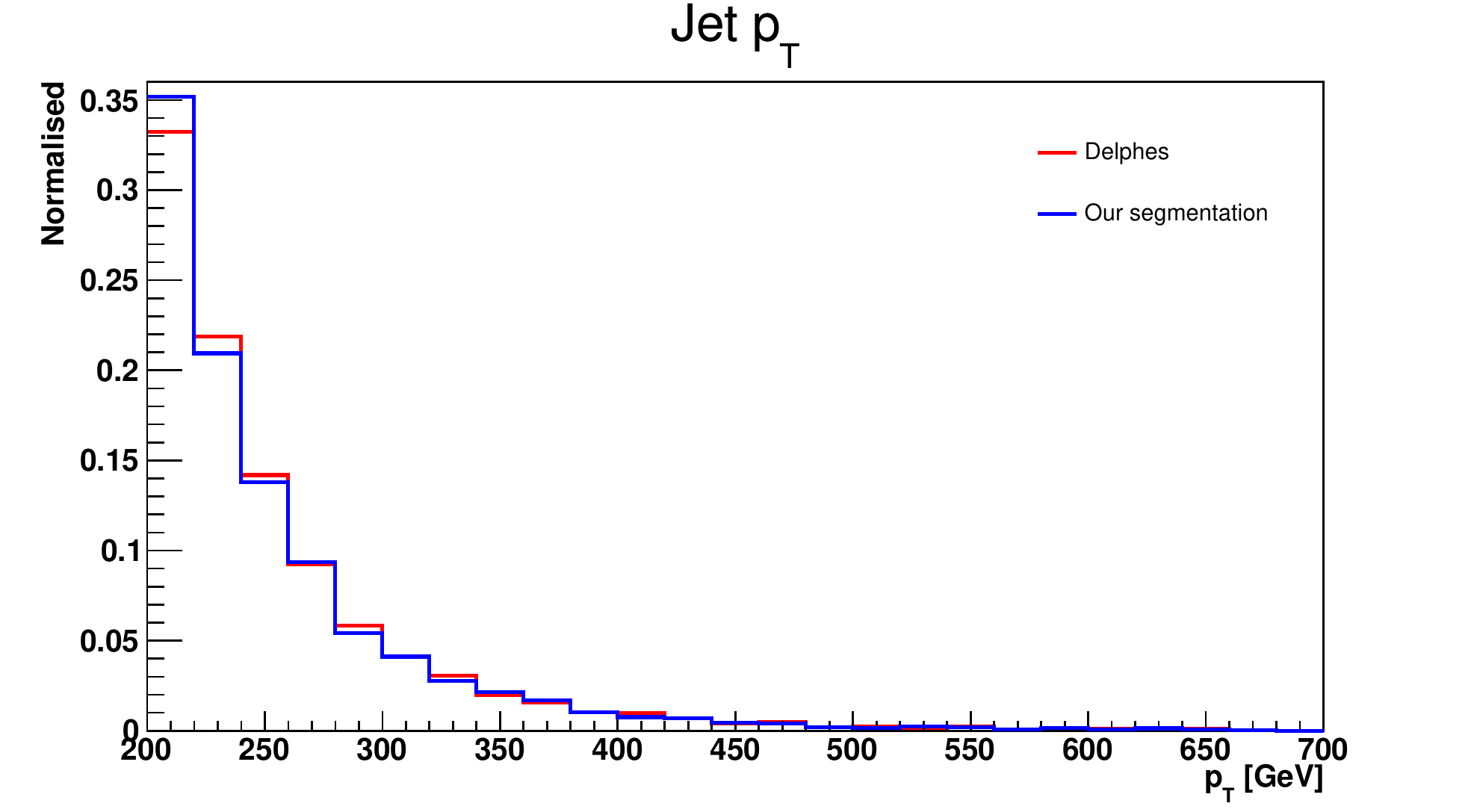}
\caption{}
\label{fig:val_pt}
\end{subfigure}%
\begin{subfigure}{0.5\textwidth}
\centering
\includegraphics[width=8cm]{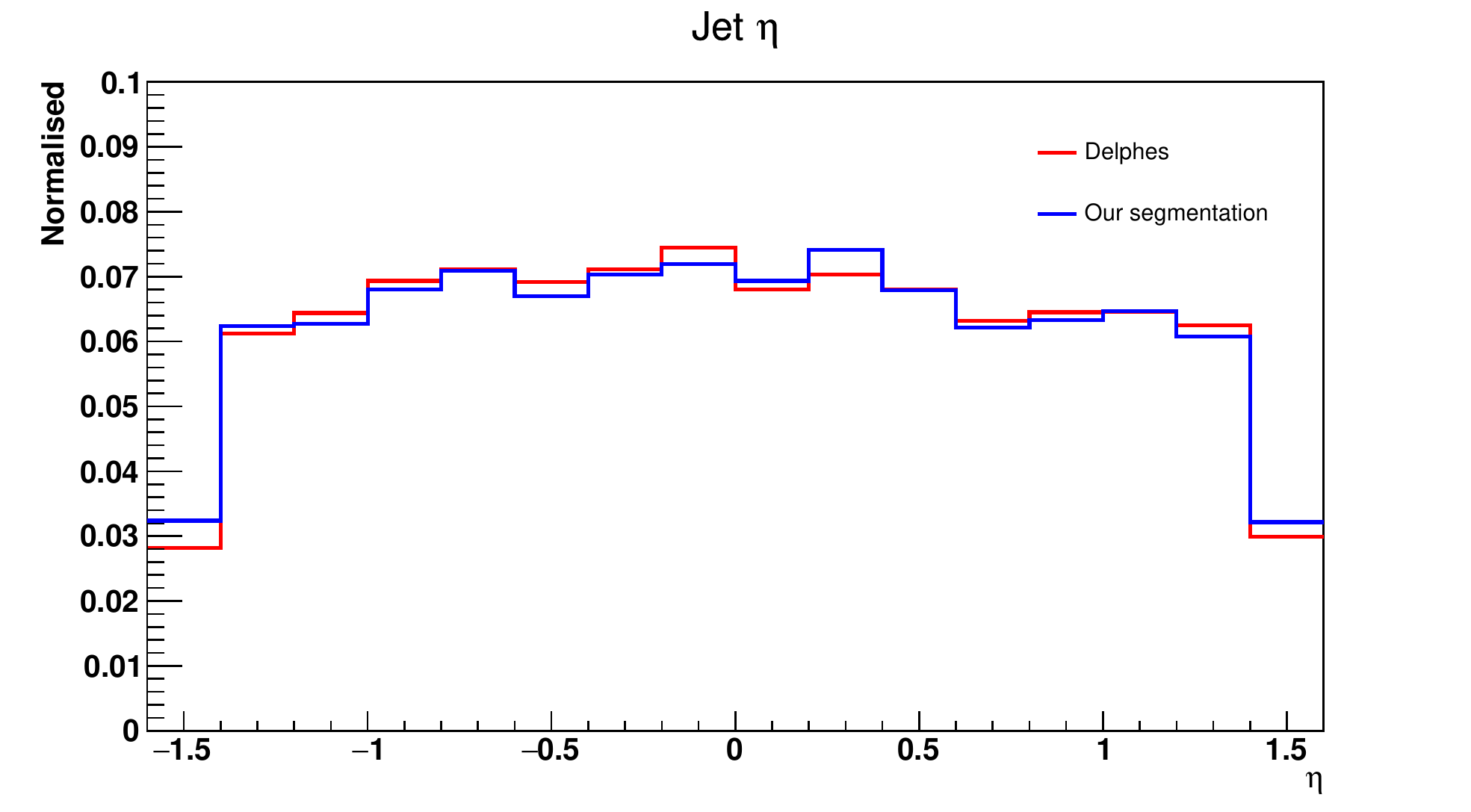}
\caption{}
\label{fig:val_eta}
\end{subfigure}
\caption{Comparison of (a) $p_T$ and (b) $\eta$ distributions using \texttt{Delphes} and our simulation for the leading jet (prompt) from a dijet sample generated with minimum $p_T$ cut of $200{\rm~GeV}$.}
\label{fig:val}
\end{figure}

Therefore, as a cross check we have made distributions of some standard processes involving prompt jets using \texttt{Delphes} as well as our segmentation and compared them. For example, we compare the $p_T$ and $\eta$ distributions for prompt jets, generated using the leading jet from a dijet sample with minimum $p_T$ cut of $200 {\rm~GeV}$ at parton level, using \texttt{Delphes} fast simulation and our own toy simulation. Fig.\ref{fig:val} shows the comparison. The shapes match well. The slight difference could be due to the fact that our layer D of the HCAL has different $\eta$ segmentation ($\Delta\eta = 0.2$) than the other layers where $\Delta\eta = 0.1$, similar to the ATLAS Tile Calorimeter, which is not present in Delphes.

Hence, our simulation is validated against Delphes, and we believe that our simulation is a good enough approximation for the ATLAS reality.

\section{Energy deposition fractions for different layers of the HCAL}
\label{app:layers}

\begin{figure}[h]
\centering
\begin{subfigure}{0.5\textwidth}
\centering
\includegraphics[width=7cm, height=4.5cm]{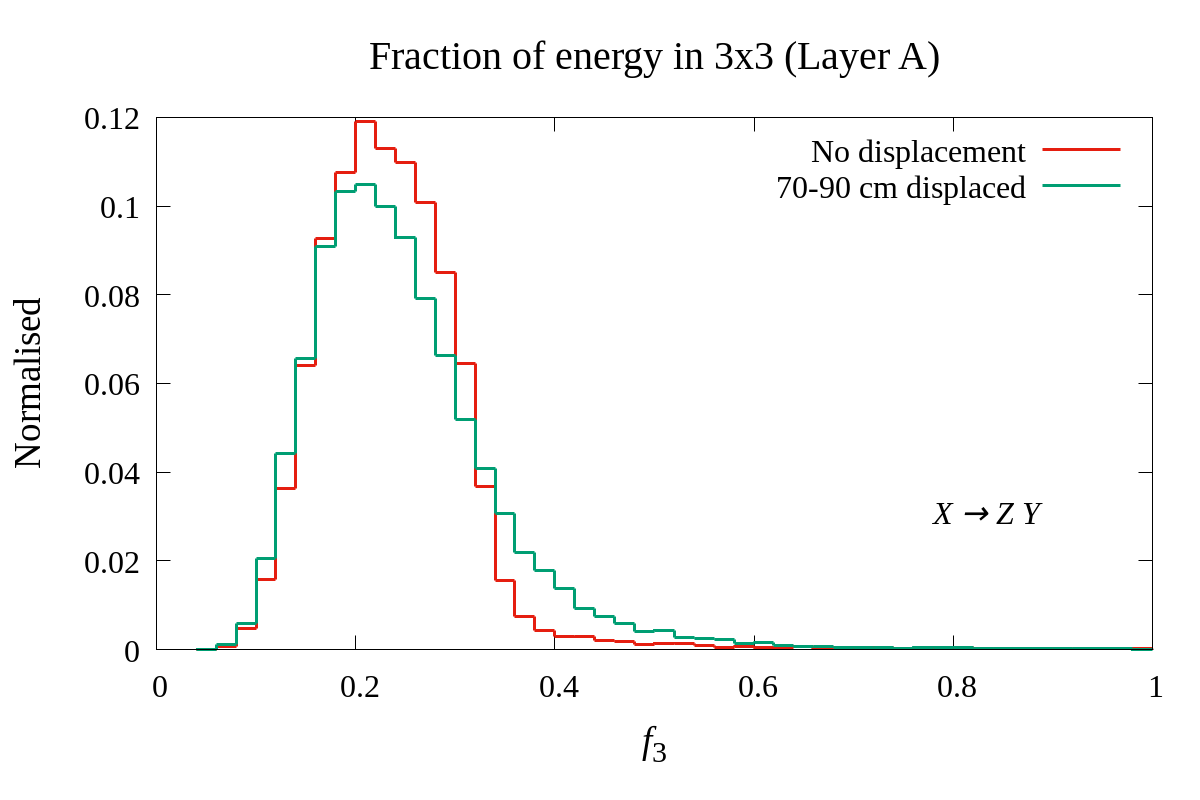}
\end{subfigure}%
\begin{subfigure}{0.5\textwidth}
\centering
\includegraphics[width=7cm, height=4.5cm]{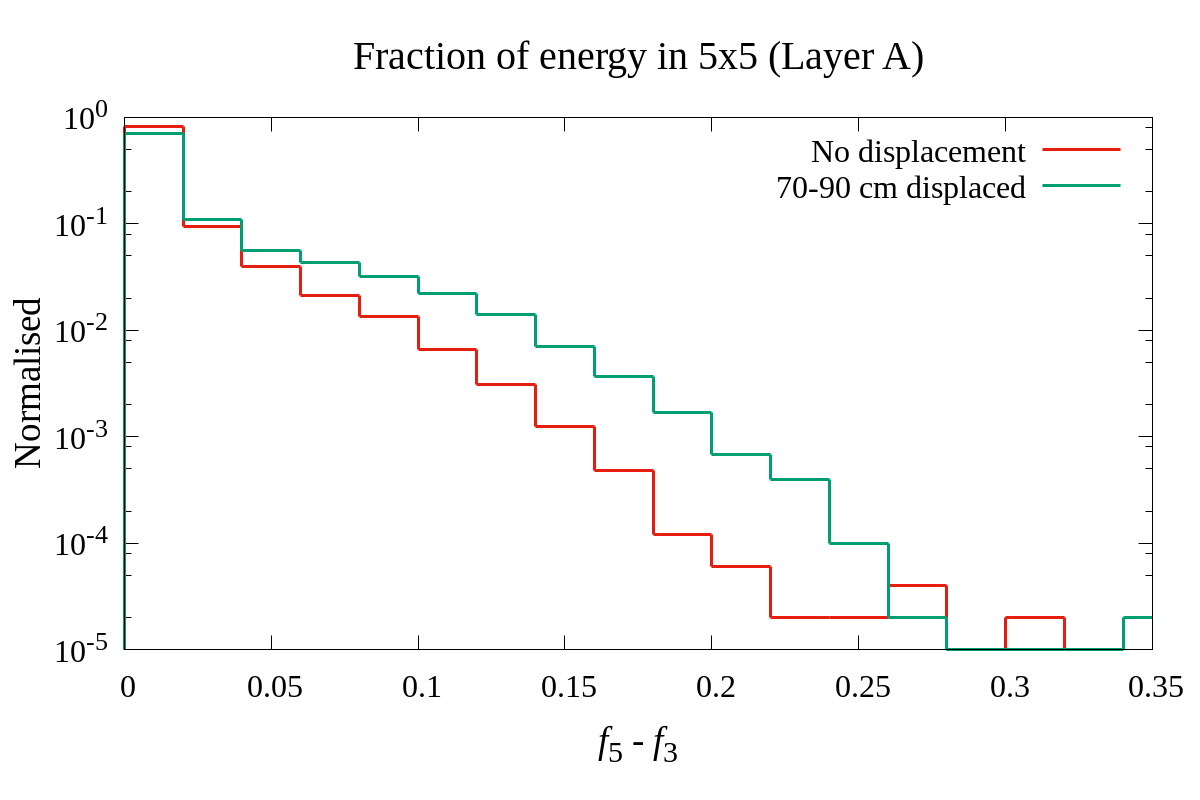}
\end{subfigure}\\
\begin{subfigure}{0.5\textwidth}
\centering
\includegraphics[width=7cm, height=4.5cm]{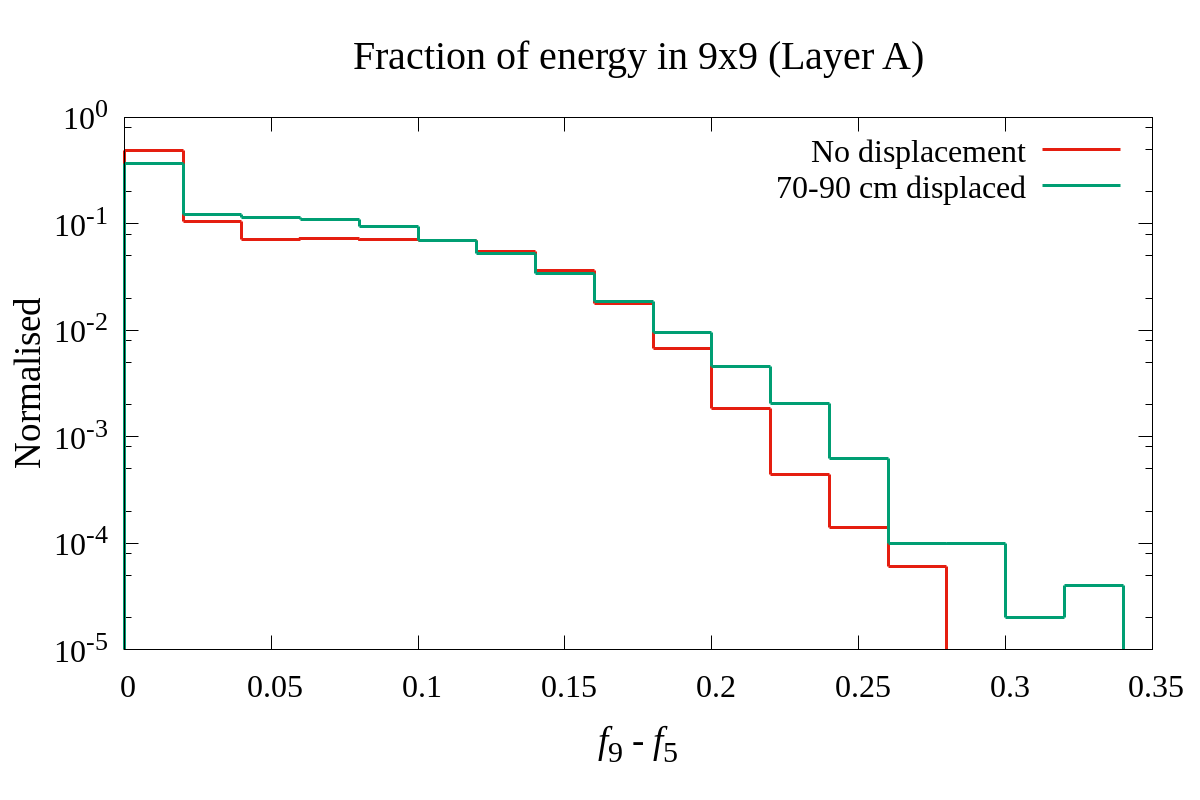}
\end{subfigure}%
\begin{subfigure}{0.5\textwidth}
\centering
\includegraphics[width=7cm, height=4.5cm]{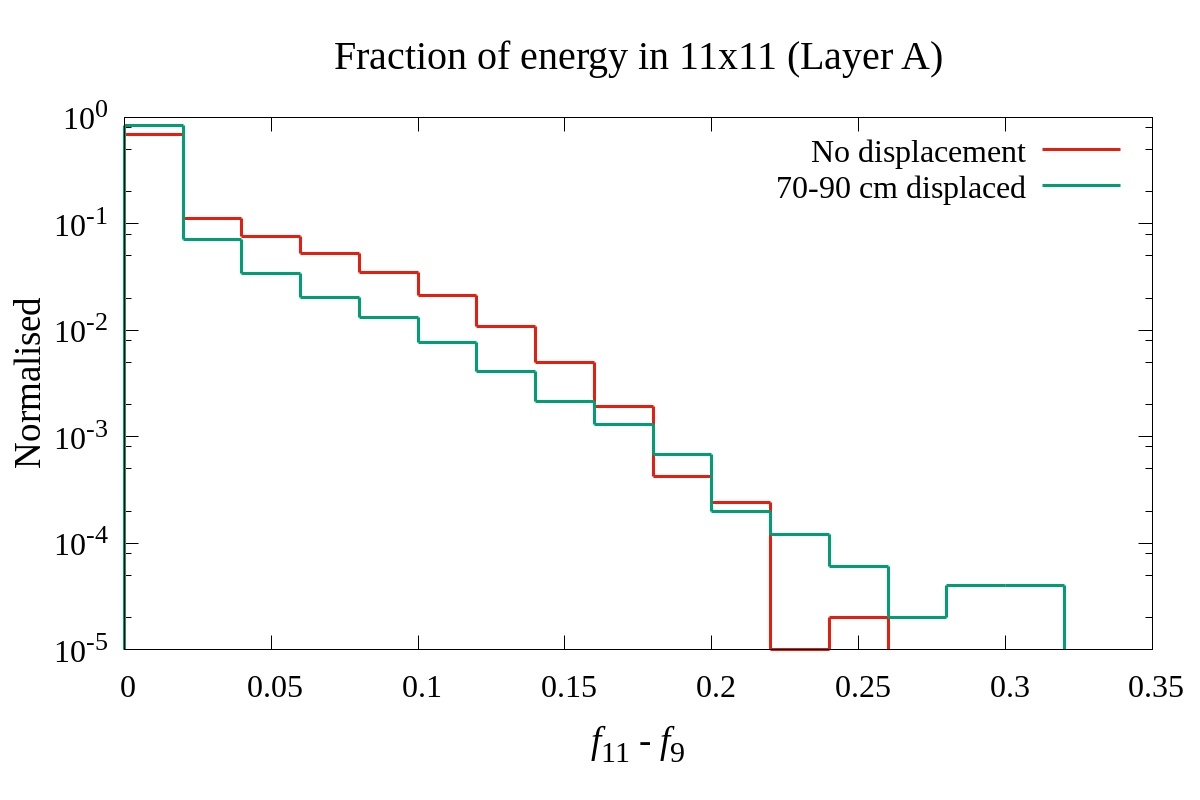}
\end{subfigure}%
\caption{Normalised distributions of energy deposition fraction ($f_i$) with varying sizes of blocks ($i\times i$) for the non-displaced and $70$-$90{\rm~cm}$ displaced $Z$ boson in layer A. 
}
\label{fig:energy_gmsbA}
\end{figure}

\begin{figure}[!htb]
\centering
\begin{subfigure}{0.5\textwidth}
\centering
\includegraphics[width=7cm, height=4.5cm]{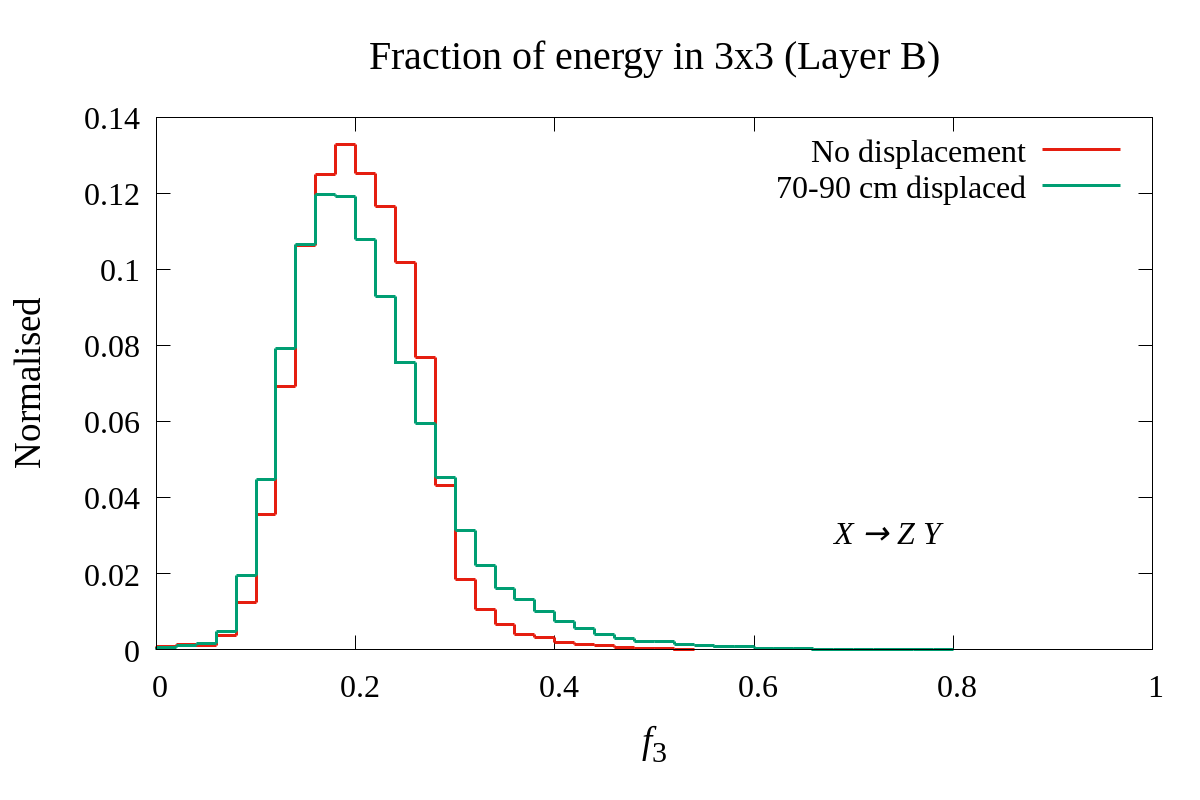}
\end{subfigure}%
\begin{subfigure}{0.5\textwidth}
\centering
\includegraphics[width=7cm, height=4.5cm]{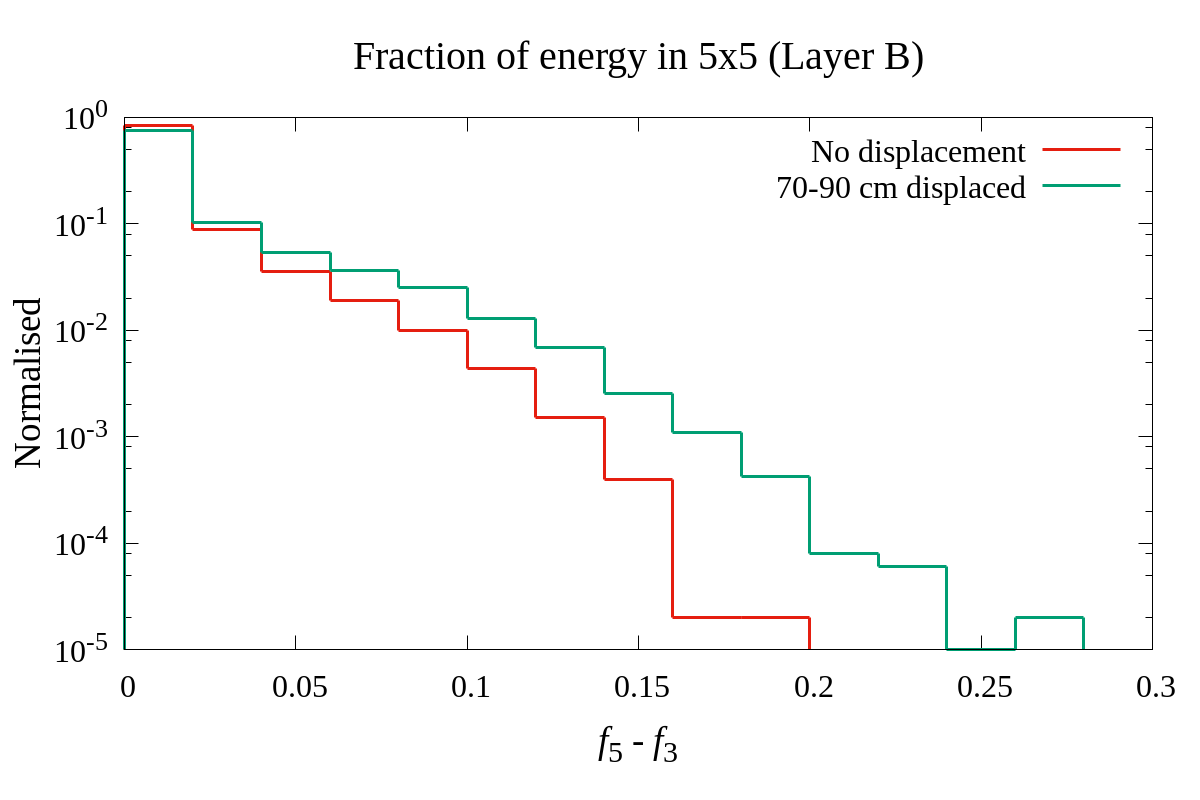}
\end{subfigure}\\
\begin{subfigure}{0.5\textwidth}
\centering
\includegraphics[width=7cm, height=4.5cm]{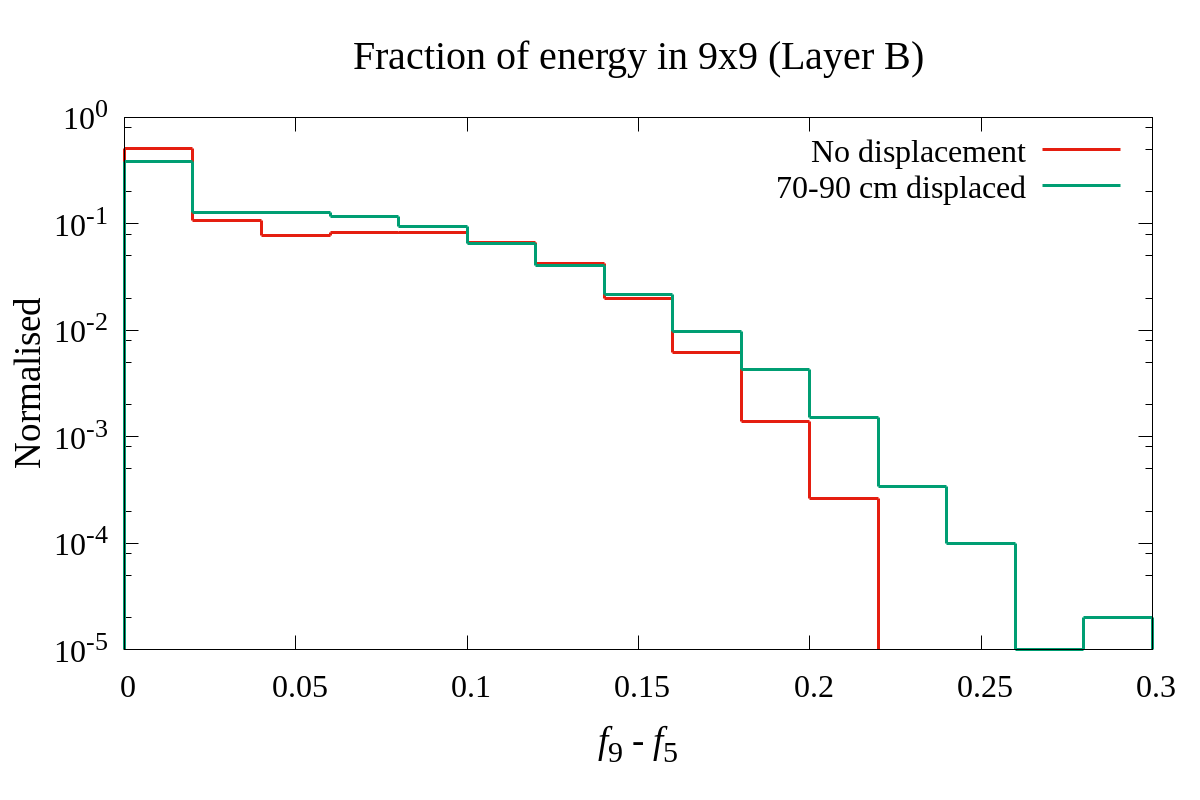}
\end{subfigure}%
\begin{subfigure}{0.5\textwidth}
\centering
\includegraphics[width=7cm, height=4.5cm]{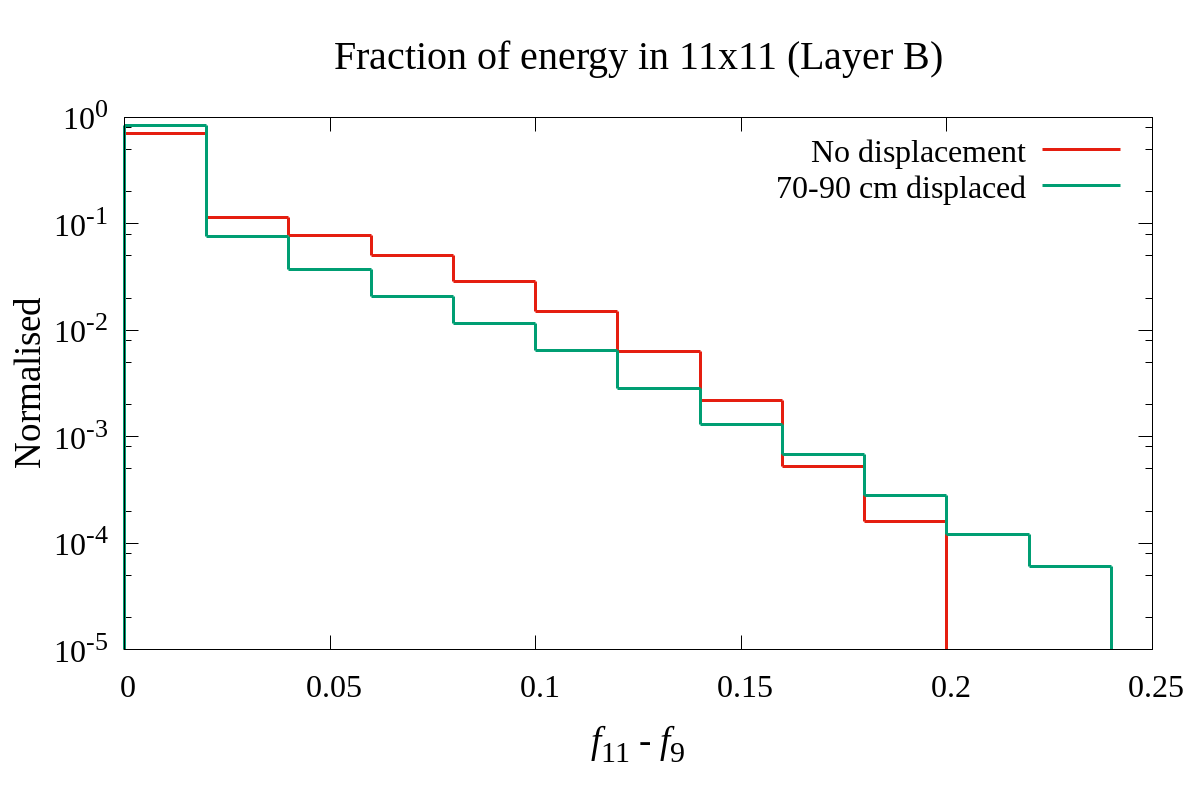}
\end{subfigure}%
\caption{Normalised distributions of energy deposition fraction ($f_i$) with varying sizes of blocks ($i\times i$) for the non-displaced and $70$-$90{\rm~cm}$ displaced $Z$ boson in layer B. 
}
\label{fig:energy_gmsbB}
\end{figure}

\begin{figure}[hbt!]
\centering
\begin{subfigure}{0.5\textwidth}
\centering
\includegraphics[width=7cm]{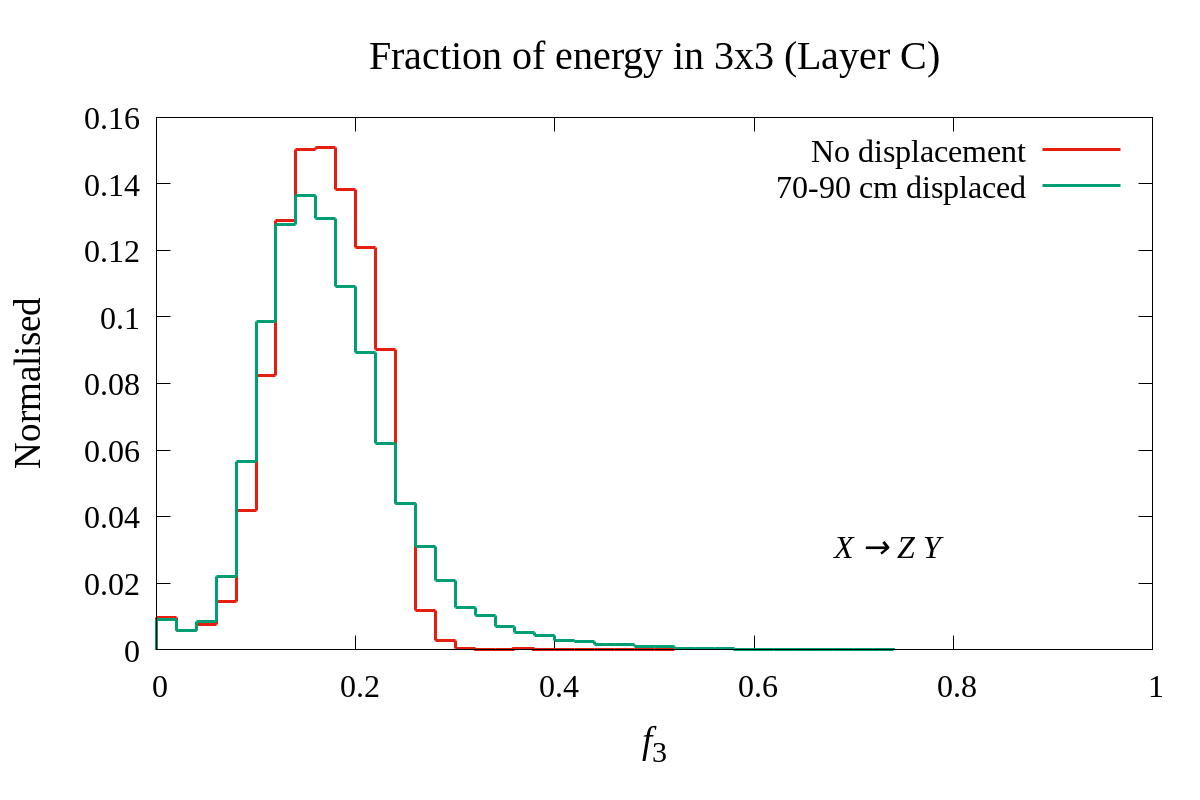}
\end{subfigure}%
\begin{subfigure}{0.5\textwidth}
\centering
\includegraphics[width=7cm]{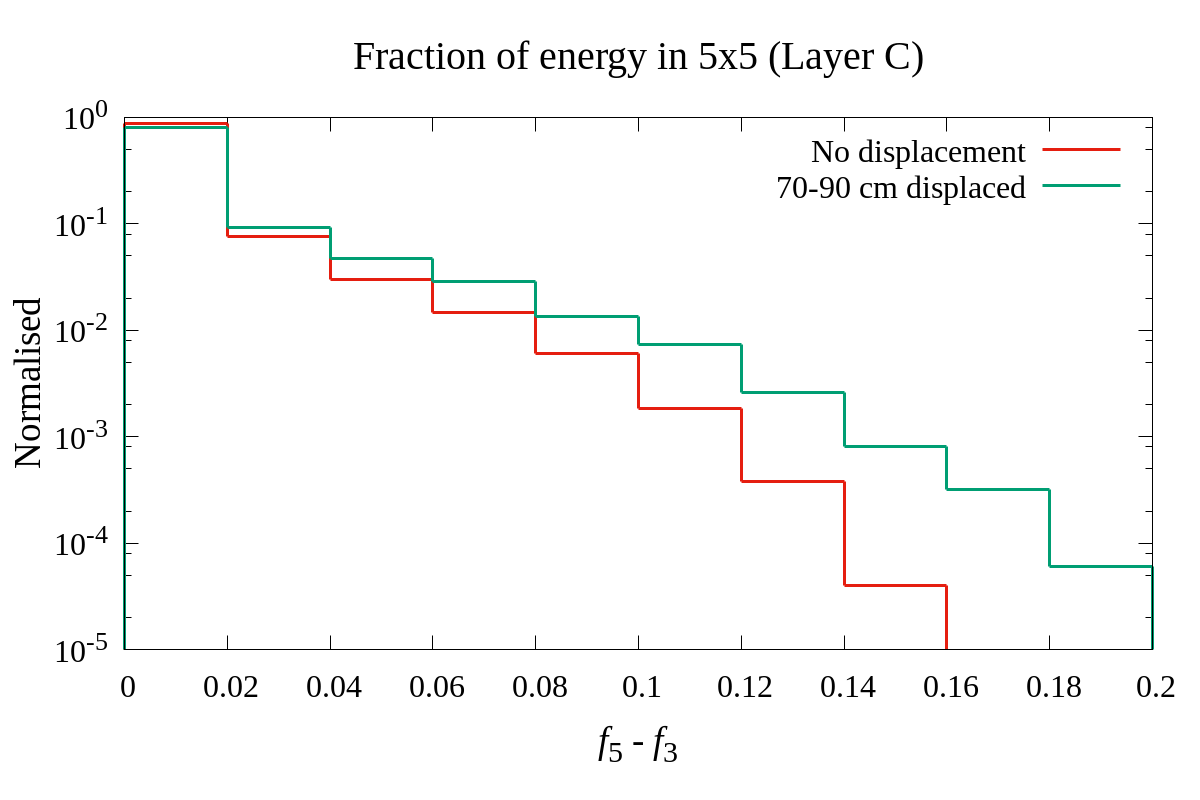}
\end{subfigure}\\
\begin{subfigure}{0.5\textwidth}
\centering
\includegraphics[width=7cm]{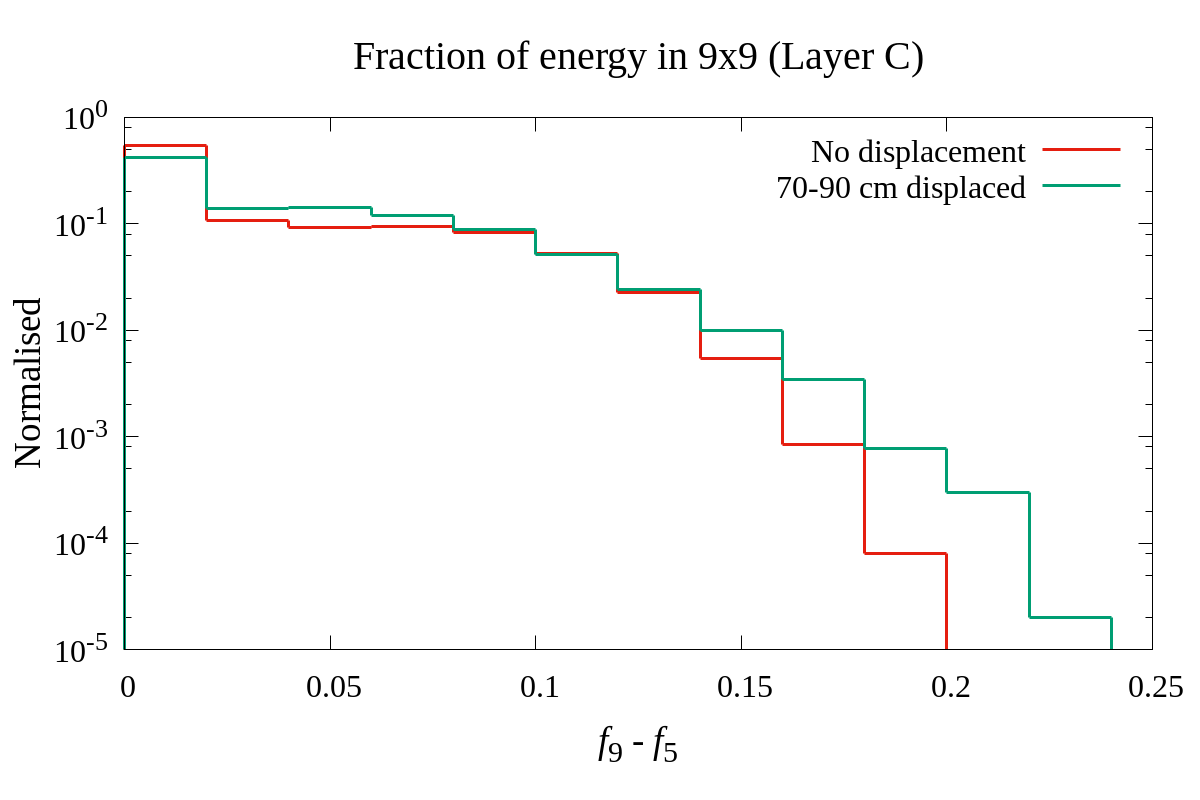}
\end{subfigure}%
\begin{subfigure}{0.5\textwidth}
\centering
\includegraphics[width=7cm]{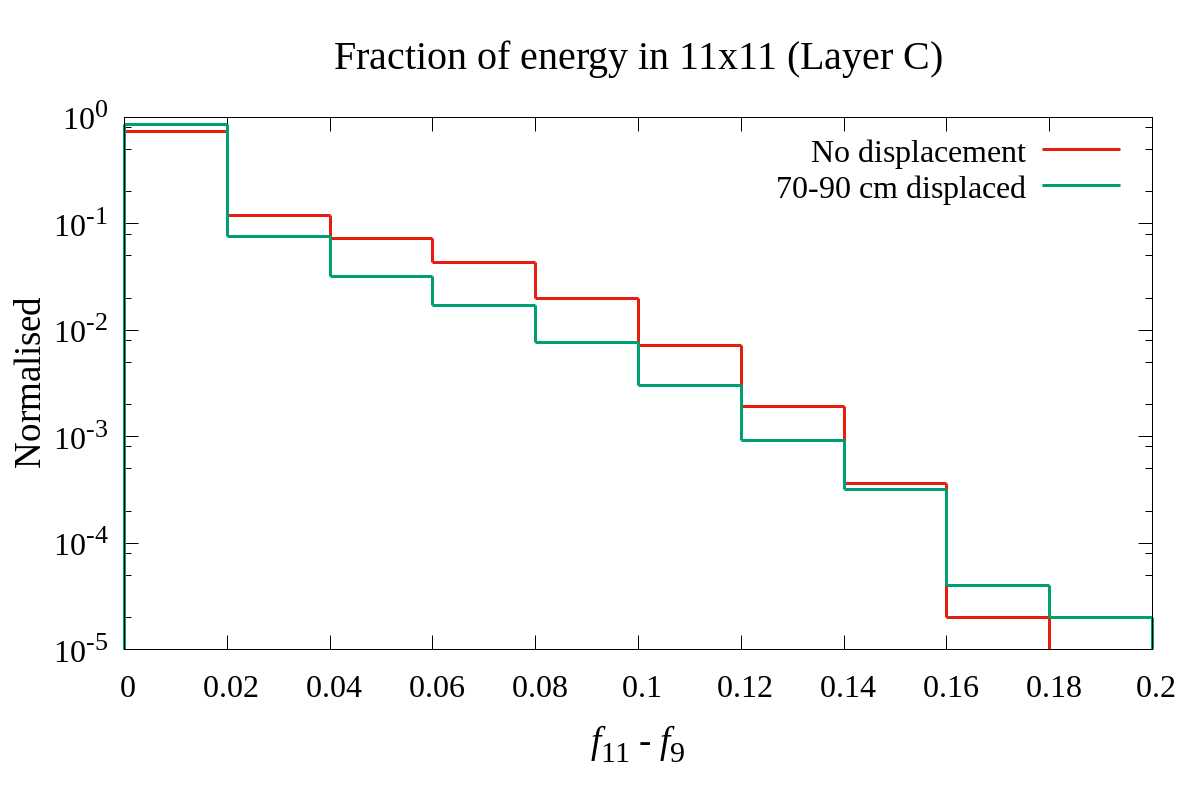}
\end{subfigure}%
\caption{Normalised distributions of energy deposition fraction ($f_i$) with varying sizes of blocks ($i\times i$) for the non-displaced and $70$-$90{\rm~cm}$ displaced $Z$ boson in layer C. 
}
\label{fig:energy_gmsbC}
\end{figure}

We here show the distributions of the energy deposition fractions for the non-displaced and $70$-$90{\rm~cm}$ displaced $Z$ boson processes for each of the starting three layers of the HCAL $-$ A (fig.\ref{fig:energy_gmsbA}), B (fig.\ref{fig:energy_gmsbB}) and C (fig.\ref{fig:energy_gmsbC}). We find a similar trend of the energy fraction in each layer as we had observed in section \ref{sssec:disp_Z} for the energy fraction taking the projection along $\eta-\phi$ of all four layers. 

In the distributions shown in figs.\ref{fig:energy_gmsbA}, \ref{fig:energy_gmsbB} and \ref{fig:energy_gmsbC}, we have calculated the fraction by taking different $i\times i$ blocks of $\eta-\phi$ regions around the highest energy bin in each layer and dividing it by the total energy deposited, after taking projection of all layers, in $28\times28$ $\eta-\phi$ region. 

In addition to this, we have also compared distributions of the fraction by taking the location of the highest energy bin after projection along $\eta$ and $\phi$ and then $i\times i$ around this bin in layer A. For the subsequent layers, we match $\eta-\phi$ of the edges the $i\times i$ region in layer A and take energy deposition within that region \footnote{For example, if the highest energy (of the final projected energy matrix) is at $i\eta=16$, for finding $f_3$, we have taken energy deposition from $i\eta$ 15 to 17 in layer A and 14 to 16 in layer B, because $\eta$ of bin 15 in layer A matches with that of bin 14 in layer B and bin 17 in layer A has same $\eta$ as bin 16 in layer B.}. Finally, the energy deposition is divided by the total energy deposited in $28\times28$ $\eta-\phi$ towers (after taking projection). We obtain similar results.

One can use information from each layer as different channels of the CNN input image as discussed in section \ref{sssec:net_arch}. In this work we are unable to do that because we are using a simplified energy deposition based on the distance of the particle from the centre of a $\eta-\phi$ bin in each layer. Our simulation lacks proper showering and interaction of the particle with detector material which can be achieved only by using full detector simulations like \texttt{GEANT}.

\end{document}